# Structure-Based Function Prediction of Functionally Unannotated Structures in the PDB

## *Prediction of ATP, GTP, Sialic Acid, Retinoic Acid and Heme-bound and -Unbound (Free) Nitric Oxide Protein Binding Sites*

**Vicente M. Reyes, Ph.D.***
**E-mail: vmrsbi.RIT.biology@gmail.com**

*work done at:

Dept. of Pharmacology, School of Medicine,
University of California, San Diego
9500 Gilman Drive, La Jolla, CA 92093-0636
&
Dept. of Biological Sciences, School of Life Sciences
Rochester Institute of Technology
One Lomb Memorial Drive, Rochester, NY 14623

**Abstract:** Due to increased activity in high-throughput structural genomics efforts around the globe, there has been a steady accumulation of experimentally solved protein 3D structures lacking functional annotation, thus creating a need for structure-based protein function assignment methods. Computational prediction of ligand binding sites (LBS) is a well-established protein function assignment method. Here we apply the specific ligand binding site detection algorithm we recently described (Reyes, V.M. & Sheth, V.N., 2011; Reyes, V.M., 2015a) to some 801 functionally unannotated experimental structures in the Protein Data Bank by screening for the binding sites of six biologically important ligands, namely: GTP in small Ras-type G-proteins, ATP in ser/thr protein kinases, sialic acid, retinoic acid, and heme-bound and unbound (free) nitric oxide. Validation of the algorithm for the GTP- and ATP-binding sites has been previously described in detail (ibid.); here, validation for the binding sites of the four other ligands shows both good specificity and sensitivity. Of the 801 structures screened, eight tested positive for GTP binding, 61 for ATP binding, 35 for sialic acid binding, 132 for retinoic acid binding, 33 for heme-bound nitric oxide binding, and 10 for free nitric oxide binding. Using the 'cutting plane' and 'tangent sphere' methods we described previously, (Reyes, V.M., 2015b), we also determined the depth of burial of the ligand binding sites detected above and compared the values with those from the respective training structures, and the degree of similarity between the two values taken as a further validation of the predicted LBSs. Applying this criterion, we were able to narrow down the predicted GTP-binding proteins to two, the ATP-binding proteins to 13, the sialic acid-binding proteins to two, the retinoic acid-binding proteins to 14, the heme-bound NO-binding proteins to four, and the unbound NO-binding proteins to one. We believe this further criterion increases the confidence level of our LBS predictions. The next logical step would be the experimental determination of the actual binding of these putative proteins to their respective ligands.

***Abbreviations:*** BS, binding site; LBS, ligand binding site; PDB, Protein Data Bank; GTP, guanosine triphosphate; ATP, adenosine triphosphate; SRGP, small Ras-type G-proteins; STPK, ser/thr protein kinase; SIA, sialic acid; REA, retinoic acid; NO, nitric oxide; hNO, heme-bound NO; fNO, free or unbound NO; PLI, protein ligand interaction(s); 3D SM, 3-dimensional search motif; CP, cutting plane; CPM, CP method; TS, tangent sphere; TSM, TS method; CPi, cutting plane index; TSi, tangent sphere index; H-bond, hydrogen bond; VDW, van der Waals; AAR, all-atom representation; DCRR, double-centroid reduced representation; Z(s), the side-chain centroid of amino acid Z; X(b), the backbone centroid of amino acid X; DCRR, double-centroid reduced representation

## 1 Introduction.

Progress in both the genomic sequencing efforts around the globe (Burley, S.K. 2000; Heinemann, U. 2000; Terwilliger, T.C., 2000; Norrvell, J.C., & Machalek, A.Z., 2000) as well as that of the various high-throughput 3D structure-determination methods (experimental or predicted) of proteins have brought about the accumulation of protein structures which completely lack functional information (Bentley et al., 2004; Murphy et al, 2004; Baxevanis, 2003; Miller et al,, 2003). For instance, the Protein Data Bank (PDB), the world's repository for protein 3D structures, has recently witnessed an accumulation of experimentally determined protein 3D structures whose functions are unknown (Berman, H.M., & Westbrook, J.D. 2004). This, in turn, has created the need for computational methods of structure-based protein function prediction, especially those which can be implemented automatically in high-throughput fashion (Jung, J.W. & Lee, W. 2004; Yakunin, A.F., et al., 2004 ). One of the main roles of bioinformatics (or computational biology) in this post-genomic era of biology is to reduce the workload of the experimentalists by computationally "eliminating" candidates for experimentation, thereby allowing them to invest their time and effort on the "good" ones that are more likely to yield useful results. This is one of the main objectives of the present work.

There are a number of established ways to predict (computationally) the function of a protein whose 3D structure (and amino acid sequence) is known. One way to do this is to predict the ligand(s) that the protein binds. To do this based on the 3D structure of the protein, one can proceed by detecting the ligand's binding site - the ligand's specific 'signature' on its receptor protein - in the receptor protein's 3D structure. Ligands usually dock on the surface of a protein, and a ligand's binding site (BS, LBS) is "buried" within the receptor protein's interior to varying degrees.

The work we describe here involves the prediction of the binding sites of six biologically important ligands, namely: GTP, ATP, sialic acid, retinoic acid and nitric oxide in heme-bound and unbound forms. The biological roles of GTP and ATP are widely established (for example, see Mazzorana M, et al., 2008, and Stork PJ., 2003, respectively) Since both GTP- and ATP-binding proteins are highly heterogeneous, we focus here on the small Ras-type G-proteins (SRGP) and the ser/thr protein kinase (STPK) families, respectively. Sialic acid (SIA) is a C9 monosaccharide, and is the key component of mucus that allows the latter to prevent infections; more importantly, however, it has a significant role in the regulation of cellular communication (Lehmann et al., 2006; Miyagi et al., 2004). Retinoic acid (REA), on the other hand, has important roles in the transcriptional modulation of certain target genes by interacting with any one of its three known receptors: alpha, beta and gamma (Germain et al., 2006; Wolf, 2006). Finally, nitric oxide (NO) is an important signaling molecule in various cell types (Cary et al., 2006; Brunori et al., 2006; Perreti et al., 2006; Russwurm et al., 2004) which may either be in heme-bound (hNO) or unbound (fNO) forms.

The binding sites of these ligands were first characterized from ligand-containing experimentally solved structures from the PDB. These collection of structures from which the binding mode of the ligands are "learned" by an algorithm is called the 'training set' for the specific ligand in question. The salient features of the binding sites are then encoded in a tetrahedral tree data structure we designate as the '3D search motif' (3D SM). Using a novel analytical screening algorithm we developed earlier (Reyes, V.M., & Sheth, V.N., 2011; Reyes, V.M., 2015a), a set of some 801 experimentally solved but functionally unannotated protein structures from the PDB were screened for these 3D search motifs. Of the 801, we detected 8 putative SRGP GTP-binding proteins, 61 putative STPK ATP-binding proteins, 35 putative SIA-binding proteins, 132 putative REA-binding proteins, 33 putative hNO-binding proteins, and 10 putative fNO-binding proteins. These candidate proteins were then subjected to the "cutting plane" and "tangent sphere" methods (Reyes, V.M., 2015b) as a further validation step. This method is a way to assess the degree of burial of a local functional site such as a ligand-binding site in a protein The validation depends on the putative structure having the same or similar depth of ligand binding site burial as those in the training structures. To our knowledge, this work is the first computational investigation that predicts the binding sites for GTP, ATP, SIA, REA, hNO and fNO from among functionally unannotated structures in the PDB, and further screens those proteins using information regarding the depth of burial of the bound ligand within its cognate receptor protein.

## 2 Datasets and Methods.

### 2.1 The Training Structures.

The screening method used here has been reported previously by us (Reyes, V.M., & Sheth, V.N., 2011; Reyes, V.M., 2015a). It requires the construction of a '3D search motif' (3D SM) from a set of training structures, and is based on the geometry and architecture of the ligand binding site (LBS) in question. The 3D SM is essentially a 'signature' of the LBS in question and contains at least six quantitative and eight qualitative parameters which are all inputted into the algorithm to enable it to detect the said LBS. The training structures for ATP-binding STPK proteins and GTP-binding SRGP proteins have been described and discussed in detail previously (ibid.). The training structures used for the construction of the 3D SM for the SIA are 1JSN, 1JSO, 1W0O, 1W0P, 1MQN (chains A and D); the training structures used to construct the 3D SM for REA are 1FM9, 1K74, 1FBY (chains A and B), 1FM6 (chains A and U), 1XDK (chains A and E), 1XLS (chains A, B, C and D), 2ACL (chains G, A, C and E); the training structures used for the construction of the 3D SM for hNO are 1OZW, 1XK3, 1ZOL (chains A and B); and finally, the training structures used to construct the 3D SM for fNO is 1ZGN, chains A and B. These training structures are all described in Table 1. The set of 801 protein structures in the PDB (all experimentally solved, mostly by x-ray crystallography) that lacked functional annotation at the time of this work are shown in Table 2. These proteins of unknown function come from many different species, but most are from *E. coli, T. maritima, T. thermophilus, B. subtilis, P. aeruginosa, H. influenzae* and *A. fulgidus*; only 18 (2.25%) come from *H. sapiens*. We used this set as the 'application set' – the set of 3D structures that we screened for the LBS's in question for the purpose of assigning function to. In addition to determining the 3D SM from the above training structures, we also determined the depth of ligand burial in each, since this information is required in the next stages of our overall screening protocol.

### 2.2 Methods

#### 2.2.1 Determination of the 3D SM's.

The overall methodology followed in this work has been described in detail (Reyes, V.M., & Sheth, V.N., 2011; Reyes, V.M., 2015a). Briefly, the set of all hydrogen bonding (H-bonding) and van der Waals (VDW) interactions between ligand and protein in the training structures are sequestered (Engh, R.A. & Huber, R., 1991); Bondi, A., 1964). Then the most dominant and/or recurrent interactions among the training structures are determined, and designated the '3D binding consensus motif'. From such a consensus interaction mode between ligand and protein, the corresponding 3D SM is constructed. The 3D SM is a tetrahedral collection of four points in space representing the protein residues most commonly in association with the ligand (in the training structures). In the 3D SM, the protein is in a reduced representation which we call the “double centroid reduced representation” (DCRR), where each amino acid is represented by two points, namely: the centroid of its backbone atoms (N, CA, C’, O), and that of its side chain atoms (CB, CG, etc.). The application set is then screened for the tetrahedral 3D SM using a screening algorithm we developed earlier (ibid.). The tetrahedral 3D SM’s for the six ligands in this study are shown in Figure 1, Panels A-D. The tetrahedral 3D SM is ***not*** just a collection of four points in space; it is a data structure that embodies a relatively large amount information about the binding site of the ligand in question. Specifically, it contains at least eight qualitative parameters, namely: the identities of the four amino acids in the tetrahedron (may be more if similar amino acids can interact with any of the ligand atoms in other receptor proteins) and their mode of association with the ligand (whether with backbone or side chain; hence, 4 x 2 = 8) and exactly six quantitative parameters (the lengths of the six sides of the tetrahedron) about the ligand binding site in question. Hence the 3D SM contains a total of 8 + 6 = 14 combined qualitative and quantitative parameters. This property makes the algorithm optimally specific for the ligand in question (ibid.).

#### 2.2.2 Determining the Degree of Burial of the Ligand Binding Sites.

In our screening protocol, there are two further steps after the detection of the LBS’s using the 3D SM method (although this step is the most crucial). These two last steps depend on the “cutting plane” and “tangent sphere” methods (CPM and TSM, respectively) of ligand burial depth quantitative determination methods we reported previously (Reyes,.V.M., 2015b; see also Figure 2). These two methods are complementary and produce numerical measures which we term the “CP index” (CPi) and “TS index” (TSi), respectively, and which are essentially quantitative measures of the degrees of burial of a given ligand or LBS. These two additional steps are meant to narrow down the set of structures testing positive for the presence of a particular LBS and thus serve to further validate the prediction results. Specifically, those which have LBS burial depths resembling those in the training structures are deemed more likely to be true positives than those whose degrees of LBS burial are quite different.

## 3 Results

The determination of the 3D SM and the validation stage (testing positive and negative control structures) for the GTP-binding site in SRGPs and the ATP-binding site in STPKs have both been presented and discussed in detail in our previous work (Reyes, V.M., 2015a), so we shall not touch upon them here and just limit our discussion in the following sections to SIA, REA, hNO and fNO binding sites.

### 3.1 Determination of the 3D SM for Sialic Acid (SIA) Binding Sites.

The H-bonds between SIA and its receptor protein in the training structures are dominated by interactions between atom N5 of SIA and the backbone O of a Gly or a Val residue in the BS; atom O1A of SIA and either an NH1 atom of an Arg or an OE1 or NE2 atom of a Gln residue in the BS; and atom O8 of SIA and the hydroxyl O of a Ser or a Tyr residue in the BS. The VDW interactions, on the other hand, are mainly between atom C7 of SIA and either the CH2 side chain atom of a Trp or the CE side chain atom of a Met residue in the BS. Careful consideration of these interactions enabled us to build the 3D search motif for SIA shown in Figure 1A.

### 3.2 Determination of the 3D SM for Retinoic Acid (REA) Binding Sites.

The H-bonds between REA and its receptor protein in the training structures are dominated by interactions between atom O1 of REA and the terminal side chain amino group of an Arg residue, and atom O2 of REA and the backbone N of an Ala residue in the BS. The VDW interactions, on the other hand, are mainly between atom C3 of REA and one of the side chain carbon atoms of an Ile or a Val residue in the BS; atom C17 of REA and either atom CB of a His or a Cys residue, or the backbone O of a Cys residue in the BS; and finally atom C20 of REA and the CD2 atom of a Phe or a Leu residue in the BS. Careful consideration of these interactions led us to construct the 3D SM for REA shown in Figure 1B.

### 3.3 Determination of the 3D SM for Heme-Bound Nitric Oxide (hNO) Binding Sites.

The H-bonds between hNO and its receptor protein in the training structures are dominated by interactions between the heme iron and the side chain amino group of a His residue in the BS; the O2D atom of heme and a side chain amino group of an Arg residue in the BS; and atom O of NO and a Gly residue atom or a side chain C atom of a Leu residue in the BS. The VDW interactions, on the other hand, are mainly between atom CHD of heme and a side chain C atom of a Phe or a Gly residue in the BS; and between atom O2D of heme and a side chain atom of a His or a Tyr residue in the BS. Careful consideration of the above interactions allowed us to build the 3D SM for hNO shown in Figure 1C.

### 3.4 Determination of the 3D SM for Free/Unbound Nitric Oxide (fNO) Binding Sites.

The H-bonds between fNO and its receptor protein in the single training structure (with two protein chains) involve N atom of fNO and the backbone N of an Arg residue or the side chain OH group of a Tyr residue in the BS. The VDW interactions, on the other hand, are mainly between the N atom of fNO and an atom of a Gly or a Val residue in the BS, or between atom O of fNO and a side chain C atom of an Ile or Phe residue in the BS. Careful consideration of these interactions allowed us to the construct the 3D SM for unbound fNO shown in Figure 1D.

### 3.5 Validation Step: Positive and Negative Controls

#### 3.5.1. Negative Control Structures.

Thirty negative control structures were used for the validation of the BS's for all six protein families studied here; they are, namely: 135L, 1A1M, 1A6T, 1BHC, 1PSN, 1BRF, 1EWK, 1CBN, 1MV5, 1JFF, 104M, 1ASH, 1B3B, 1BRF, 1CKO, 1CRP, 1EWK, 1F3O, 1FW5, 1HWY, 1JBP, 1MJJ, 1MV5, 1NQT, 1OGU, 1PE6, 1RDQ, 1SVS, 1TWY and 1Z3C. The above structures are all described in Table 3. Our results show that in all cases, the algorithm found no 3D SM in any of the negative control structures as expected. These results imply that the algorithm is highly specific for their respective ligands.

#### 3.5.2. Positive Control Structures.

As for positive control structures, we note that there are no other appropriate positive structures in the PDB for the four above ligands as all of them have been used as training structures. Positive control structures to be used for validation must be yet "unseen" by the algorithm. We thus constructed artificial positive control structures from the negative control structures by replacing four appropriate amino acid residues in the latter to make a legitimate 3D SM for the particular ligand. These artificially mutated structures were then screened for the appropriate 3D SM using our algorithm. In all cases, the algorithm detected the artificially embedded 3D SM for the particular ligand (data not shown). These results imply that the screening algorithm has high sensitivity for the 3D SM corresponding to the particular ligand.

### 3.6 Screening Results

The screening process is illustrated in Figure 2. There are three stages in our screening process, the first stage and the most important being the LBS determination. The next stages involve the determination of the LBS burial in the putative structures from the preceding stage. This is done by determining their CPi and TSi, respectively. The computed values are compared against the CPi and TSi of the respective training structures, and those putative structures having CPi and TSi closest to any of those of the training structures are deemed "double positives", and are thus are considered best ligand-binding candidates in their respective protein families (see below). The application set, the 801 functionally unannotated structures in the PDB that served as application structures for this study, is shown in Table 2. These proteins come from a diverse distribution of species (see Table 4). The 'Cutting Plane' and 'Tangent Sphere' methods, on the other hand, are illustrated in Figure 1, Panels A and B, of our previous paper (Reyes, V.M., 2015b), which schematically illustrate the two methods and how they complement each other.

Overall results are as follows: of the 801 application structures, we detected 61 putative ATP-binding STPK proteins (7.6%), eight GTP-binding SRGP proteins (1.0%), 35 putative SIA-binding proteins (4.4%), 132 putative REA-binding proteins (16.5%), 33 putative hNO binding proteins (4.1%), and 10 fNO binding proteins (1.2%). We now show the details of these screening results in the following sections. Note that a protein that tested positive for a particular LBS may have more than one chain, and one or more LBSs may have been detected in each chain.

In the first 6 subtables of Table 5, the blue entries on top are the training structures for the particular 3D SM. Meanwhile, the black entries below are the structures that tested positive for the ligand in question. The headings "CPM" and "TSM" stand for "cutting plane" and "tangent sphere" methods, respectively. The red arrows point out those positive structures whose CPM and TSM indices are either within an arbitrarily set difference, e.g., within 8-10%, from any one of those of the training structures, respectively, of the closest one in the set. In each case, integration of these ligand burial depth results with those of the LBS screening results further trim down the positive set, at the same time further validating the LBS existence prediction. The information contained in the different parts of the tables are illustrated and explained diagrammatically in part 7 (of 7) of Table 5. Note that due to the large number of structures testing positive for the LBS (first stage of screening) in question in the two cases of ATP-binding STPK and REA-binding protein families (Table 5, part 2 of 7 and part 4 of 7, respectively) this diagram is not strictly adhered to. Instead, only structures with CPi and TSi values within 10.0 Å of those of a training structure are shown.

**3.6.1 Screening Results for GTP-Binding Sites in Small Ras-type G-Proteins.** Eight structures (1.0% of the original 801) tested positive in the initial screening step, the detection of the 3D SM for GTP (Table 5, part 1 of 7). This set then got reduced to seven (0.9%) after matching their CPi or TSi (i.e., at least one of them) values to those of the training structures. From these seven structures, two (0.2%) stand out, namely, 1XT1 and 1RU8, because ***both*** of their CPi and TSi values are close to those of one of the structures in the training set (see Table 6).

**3.6.2 Screening Results for ATP-Binding Sites in ser/thr Proein Kinases.** The number of structures that tested positive for the ATP BS for this family is 61 (7.6%; see Table 5, part 2 of 7). By incorporating the ligand burial depth data from the CP and TS methods, 24 of the 61 structures testing positive for the ATP-binding site have been eliminated, leaving 37 structures (4.6%). Out of these 37, the following 11 to 13 structures (ca. 1.6%) are strong candidates because their CPi's and TSi's resemble both those of a training structure: 1WM6, 2CV1, 1RKQ, 1NF2, 1TQ6, 1MWW, 1TT7, 1T57, 1F19, 1RKI, 1Y9E (and possibly 1VPH and 1YYV as well; see Schwarzenbacher R. et al., 2004; Teplyakov A. et al., 2002; Beeby M. et al., 2005; Kunishima N. et al., 2005; see also Table 6).

**3.6.3 Screening Results for Sialic Acid Binding Sites.** For this ligand, 35 (4.4%) structures tested positive for the SIA binding site (Table 5, part 3 of 7). Of these, only 20 (2.5%) possess either a CPi or TSi close to that of a training structure. Of these 20, two structures (0.2%) namely, 1VKA and 1IUK, stand out as both of their CPi and TSi values resemble both the CPi and TSi values of one of the structures in the training set for this ligand (see Table 6).

**3.6.4 Screening Results for Retinoic Acid Binding Sites.** Of the 801 application structures, 132 (16.5%) tested positive for the REA binding site. Incorporating the ligand burial depth data from the CPM and TSM methods, almost 60% of the above 132 structures have been eliminated, leaving 53 candidate structures (6.6%; Table 5, part 4 of 7). The following 13 or 14 structures (ca. 1.7%) are strong candidates because both their CPi's and TSi's resemble both those of a training structure: 1YEY, 1NX4, 1TU1, 1Y8T, 2EVR, 1WU8, 1R1H, 1U61, 1T6S, 1NX8, 1NJH, 1Z6M, and 1VIM (and possibly 1ZE0 as well; see Clifton, I.J. et al., 2003; Asch WS, Schechter N., 2000; see also Table 6).

**3.6.5 Screening for Heme-Bound NO Binding Sites**. For this ligand, 33 structures (4.1%) tested positive for the hNO binding site. They have been further trimmed down to 12 (1.5%) upon incorporation of the ligand burial data using the CPM and TSM (Table 5, part 5 of 7). Of these 12, four structures (0.5%), namely 1ZSW, 1VKH, 1UAN and 2B4W stand out as their CPi and TSi values resemble both those from a training structure for this ligand (see Arndt, J.W. et al. 2005; Zhou C.Z. et al., 2005; see also Table 6).

**3.6.6 Screening for Free/Unbound NO Binding Sites.** In this set, the 10 structures (1.2%) tested positive for the fNO binding site. These have been narrowed down to six (0.7%) upon including the results from the CPM and TSM ligand burial data (Table 5, part 6 of 7). Of these six, a single structure (0.1%), namely, 1UC2, stands out as its CPi and TSi values both resemble those by the lone training structure, 1ZGN., for this ligand (see Table 6).

## 4 Discussion.

Using a novel analytical screening algorithm we developed earlier (Reyes, V.M., & Sheth, V.N., 2011; Reyes, V.M., 2015a), we have screened some 801 functionally unannotated x-ray diffraction structures deposited in the PDB for the binding sites of GTP, ATP, sialic acid, retinoic acid, and heme-bound and unbound nitric oxide. We detected eight SRGP GTP-binding sites, 61 STPK ATP-binding sites, 35 SIA-binding sites, 132 REA-binding sites, 33 hNO-binding sites and 10 fNO binding sites, with some structures containing more than one binding site for the ligand in question. The detection of the LBS for a particular ligand was accomplished by detecting the 3D SM for that ligand in the protein structures. This idea depends on the assumption that the 3D SM (and hence the binding site characteristics) for a given ligand is conserved within a protein family.

Using another novel analytical method we developed earlier (Reyes, V.M., 2015b) called the "cutting plane" and "tangent sphere" methods, the degrees of burial of these ligand binding sites were also determined and used as a further validation step for the ligand binding prediction. Thus the positive structures above were further culled by comparing their CPi or TSi to those of the training structures for the protein family and those which had similar values were retained, the rationale being those which have depths of LBS burial resembling those in the training structures are deemed more likely to be true positives than those who do not. This criterion depends on the reasonable premise that ligand burial depth is characteristic of a particular ligand-binding protein family.

Our LBS detection method depends on the availability of protein complex 3D structures with the bound ligand under study and as such relies heavily on the contents of the PDB. Although experimental structures for GTP- and ATP-binding proteins abound in the PDB, structures of proteins bound with other ligands are underrepresented. For example, the scarcity of structures containing SIA, REA, hNO and fNO in the PDB is a limitation in terms of having an ample number of both training and control (validation) sets for our screening algorithm. However, since our screening algorithm is largely analytical, the need for exhaustive positive and negative control structures is not that critical compared to statistical algorithms such as those based on SVM and neural networks. This is one advantage of an analytical algorithm over a stochastic one.

The fuzzy factor or margin, ε, we incorporate into the branches and node-edges in the 3D SM are usually in the order of 1.0 - 1.5 Å (Reyes, V.M. & Sheth, V.N., 2011; Reyes, V.M., 2015a). Thus in cases where the protein assumes drastic conformational changes upon ligand binding and displacements of amino acid residues at the binding site are much greater than 1.5 Å, our method will perhaps likely fail. We believe it is reasonable to assume that the deeper within the protein interior the LBS lies, the more drastic the conformational changes the protein undergoes upon binding the ligand (i.e., in transitioning from the 'apo' to the complexed form). But whether or not the predictive power of our algorithm decreases as the LBS lies deeper within the protein remains to be investigated.

In the determination of H-bonds between protein and ligand to build the 3D SM, we did not ascertain the linearity of the bonds of the interacting atoms between ligand and protein (amino acids in the BS); we merely measured non-hydrogen interatomic distances and we sequester only those with perfect or near-perfect H-bond distances (2.7Å-2.9 Å). Thus this issue is unlikely to have a significant adverse effect on our results, as instances in which the H-bonding atoms have perfect or near-prefect H-bonding distances and at the same time non-linear, are quite rare.

## 5 Summary and Conclusions.

By determining the most prevalent and/or dominant H-bonding and VDW interactions between ligand atoms and amino acid residue atoms in the BS of its receptor protein, we have constructed a 'signature' of the binding sites of six biologically important ligands – GTP, ATP, SIA, REA, hNO and fNO. We designate this 'signature' as the 3D BS consensus motif for the particular ligand. We have then encoded these binding site signatures in a tetrahedral tree data structure we call the 3D search motif or "3D SM" for the ligand in question. Then, using a novel analytical search algorithm we developed earlier (Reyes, V.M., & Sheth, V.N., 2011; Reyes, V.M., 2015a) experimentally determined protein structures in the PDB that lacked functional annotation were screened for the above five ligands. We detected eight structures with the GTP-binding site of the SRGP family, 61 structures with the ATP-binding site of the STPK family, 35 structures with the SIA binding site signature, 132 with the REA's, 33 with the heme-bound NO's, and 10 with the free NO's. The positive proteins above were further subjected to validation by determining the depth of burial of their LBS's using their CPi and TSi values and comparing them to those of their training structures. Respectively seven, 37, 20, 53, 12, and six of the GTP-, ATP-, SIA-, REA-, hNO- and fNO-binding proteins had ***either*** their CPi ***or*** TSi close to those of a retaining structure for the protein family. Of these, respectively two, 28 (of which 13 stand out from the rest), two, 30 (of which 14 stand out from the rest), four and one of the GTP-, ATP-, SIA-, REA-, hNO- and fNO-binding proteins had ***both*** of their CPi ***and*** TSi close to those of a retaining structure for the protein family. Thus by incorporating information about the depth of LBS burial in the positive proteins from the 3D SM screening, they can be further narrowed down significantly for increased confidence in the LBS prediction. At this point in the protein function prediction process, the job of the bioinformaticist is usually done and the experimentalists take over. Thus we are currently awaiting experimental verification of the results we report here. Our final results are shown in Table 6.

**Acknowledgments.** This work was supported by an Institutional Research and Academic Career Development Award to the author, NIGMS/NIH grant number GM 68524. The author also wishes to acknowledge the San Diego Supercomputer Center, the UCSD Academic Computing Services, and the UCSD Biomedical Library, for the help and support of their staff and personnel. He also acknowledges the Division of Research Computing at RIT, and computing resources from the Dept. of Biological Sciences, College of Science, at RIT.

**FIGURE LEGENDS:**

**Figure 1, Panels A-D: The 3D Search Motifs.** The 3D search motifs for the four ligands under study are shown: Panels A-D: Sialic acid, retinoic acid, heme-bound nitric oxide, and unbound nitric oxide search motifs, respectively. The lengths of the six sides of the tetrahedral motif (in Å) are also shown in an accompanying side table; the numbers inside parentheses are the corresponding standard deviations from the training structures. The ligand in each case is shown with its component atom names. The amino acids representing the tetrahedral vertices are indicated, with "(s)" indicating side chain interaction with ligand, and "(b)", backbone interaction. The root and three nodes are also indicated by the boxed letters.

**Figure 2. The Elimination Process.** Both local and global structure information are utilized in the process of elimination to search for candidate positive structures. Set A, the outermost red circle, represents the starting test/application set composed of 801 PDB structures without functional annotation. They are first screened for the particular 3D SM in question, and those that test positive, i.e., those that possess the 3D SM, form a subset of A; we call it set B (blue circle). Set B structures are then subjected to the "Cutting Plane" and "Tangent sphere" Methods (Reyes, V.M., 2015b). The CPM and TSM indices (CPMi and TSMi, respectively) of each structure are then compared respectively to those of t he training structures used to create the 3D SM's. Those whose CPMi or TSMi are within several units (typically 8-10) of those of the training structures, are considered to have similar indices, and form a subset of B; we call it set C (brown circle). Structures in set C are further analyzed to determine whether their indices are both respectively similar to those any one or more of the training structures. Those which satisfy this criterion form a subset of C, which we call set D (green circle). This main advantage of this elimination procedure is it can be automated and ran in batch or high-throughput mode, without the requirement for human intervention, a feature desired of analytical tools for large datasets.

**TABLE LEGENDS:**

**Table 1. The Training Sets**. The training structures for the determination of the 3D SMs for the biding sites of sialic acid, retinoic, and heme-bound and unbound nitric oxide, are shown. The PDB IDs of the structures are shown on column 1, the source organism on column 2, and a brief description of the structures is on column 3.

**Table 2. The Control Structures.** Negative control structures for SA, RA, hNO and fNO binding sites are shown. As for positive control structures for those ligand binding sites, please see text. Positive and negative control structures used for validating the 3D SM's for GTP-binding SRGP and ATP-binding STPK protein families are taken up in detail in our previous work (Reyes, V.M., 2015a).

**Table 3. The 801 Functionally Unnanotated Proteins in the PDB Used As Application Structures.** These 801 structures were obtained from the PDB in early 2006 by querying the PDB search site with the words "unknown function" or similar phrase. The absence of functional annotation in all 801 structures was further confirmed by examining the header information in each PDB file, which contained the phrase "function unknown" or a similar one in each case.

**Table 4. Species Distribution of Application Set.** The species distribution of the 801 application structures is shown in this table. The 801 structures come from 104 known species (that include bacteria, archaea, protozoans, and some higher organisms including humans), an uncultured bacterium (unknown species), and one is a synthetic protein. The 5 most represented species are *E. coli* (11.0%), *T. maritima* (7.9%), *T. thermophilus* (7.0%), *B. subtilis* (6.0%) and *P. aeruginosa* (4.7%).

**Table 5, Parts 1-7. Cutting Plane and Tangent Sphere Indices Used to Assess LBS Burial in Screening Results: GTP in Small Ras-type GP (part 1 of 7); ATP in ser/thr PK; (part 2 of 7); Sialic Acid (part 3 of 7); Retinoic Acid (part 4 of 7); Heme-Bound NO (part 5 of 7); Unbound NO (part 6 of 7).** The information in the above six tables is illustrated and identified schematically in part 7 of 7 of the table. Results of the determination of the particular binding site burial using the "cutting plane" and tangent sphere" methods are shown (headings "CPM" and "TSM", respectively). The degree of burial is expressed as % of protein atoms on the exterior side of the cutting plane and inside the tangent sphere, respectively. Part 7 of 7 diagrammatically identifies what information are contained in the tables above based on their location in the table.

**Table 6. Application Structures that Tested Positive.** The structures from the set of 801 functionally unannotated proteins (Table 3) in the PDB that tested positive of the 3D SMs of GTP (in small, Ras-type G-proteins, ATP (in ser/thr protein kinases), sialic acid, retinoic acid, and heme-bound and unbound nitric oxide are summarized in this table. Note that most of the structures are still functionally unannotated at the time of this writing, as shown by the scarcity of entries in the last column, which is the published reference papers for the particular structure (see also References section).

# FIGURES:

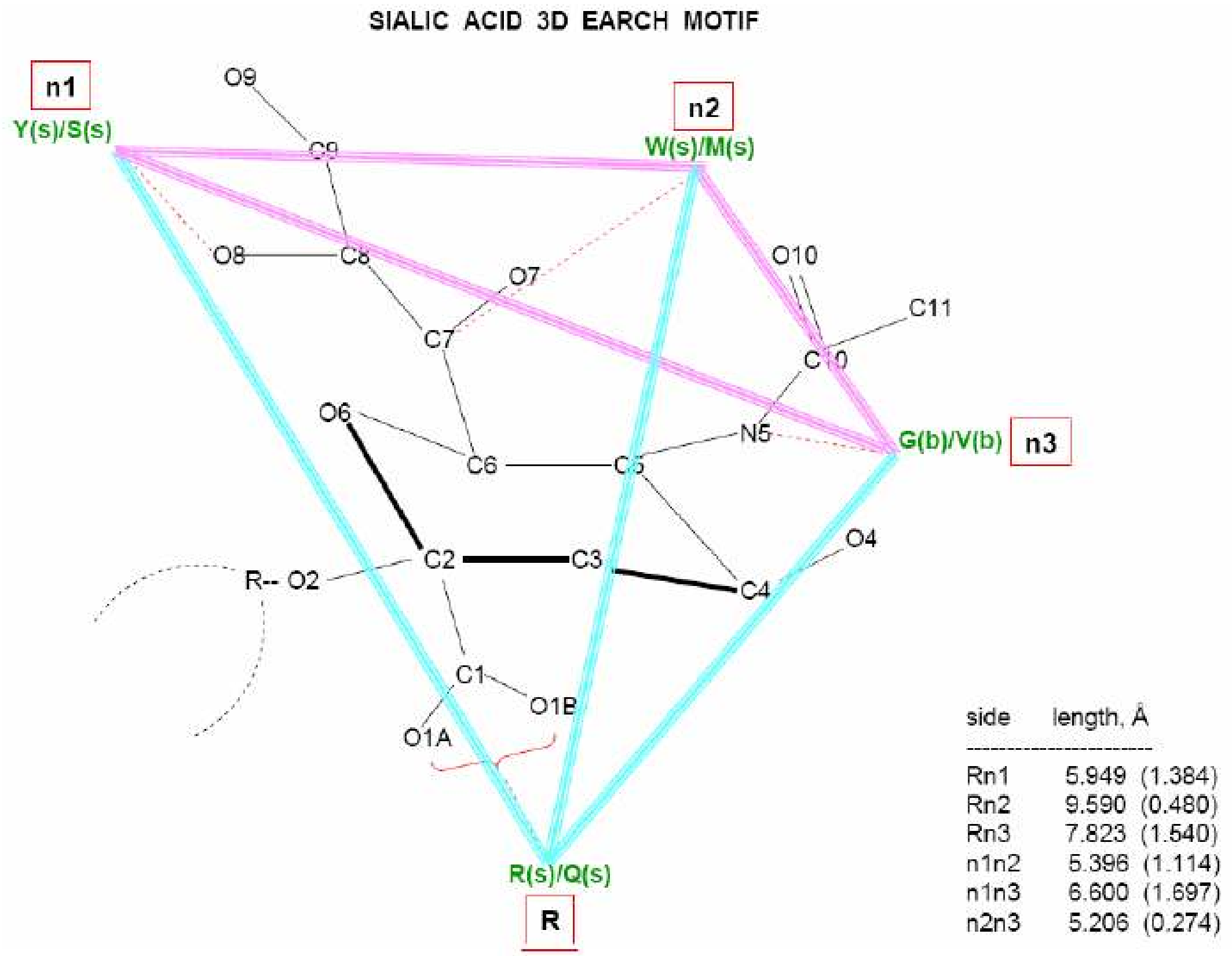

| side | length, Å |
|---|---|
| Rn1 | 5.949 (1.384) |
| Rn2 | 9.590 (0.480) |
| Rn3 | 7.823 (1.540) |
| n1n2 | 5.396 (1.114) |
| n1n3 | 6.600 (1.697) |
| n2n3 | 5.206 (0.274) |

**Figure 1A.**

**RETINOIC ACID 3D SEARCH MOTIF**

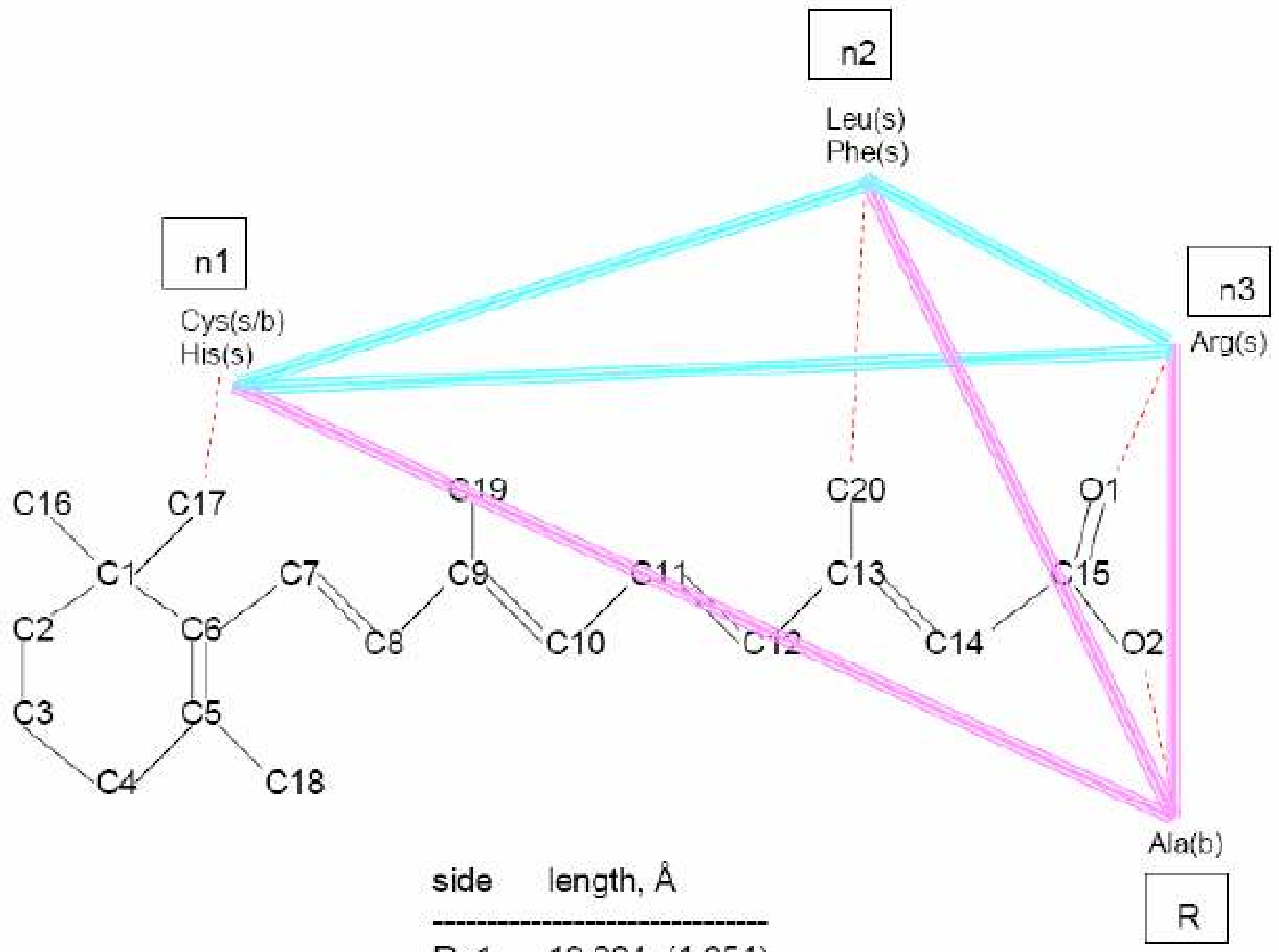

| side | length, Å |
|---|---|
| Rn1 | 19.224 (1.654) |
| Rn2 | 7.869 (1.784) |
| Rn3 | 5.622 (0.120) |
| n1n2 | 13.008 (2.710) |
| n1n3 | 18.819 (1.884) |
| n2n3 | 7.402 (0.326) |

**Figure 1B.**

## 3D Search Motif #1: Heme-bound Nitric Oxide:

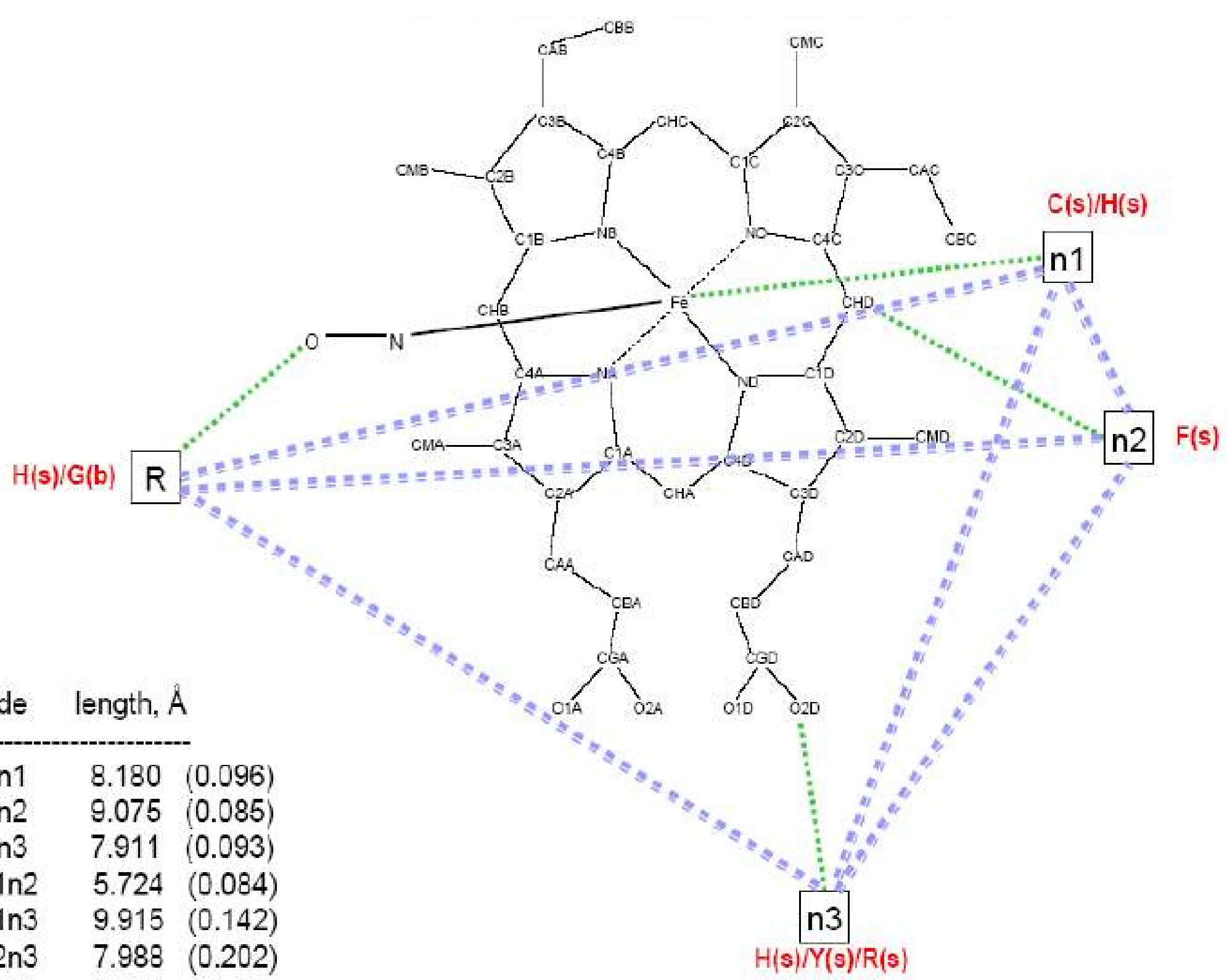

| side | length, Å | |
|---|---|---|
| Rn1 | 8.180 | (0.096) |
| Rn2 | 9.075 | (0.085) |
| Rn3 | 7.911 | (0.093) |
| n1n2 | 5.724 | (0.084) |
| n1n3 | 9.915 | (0.142) |
| n2n3 | 7.988 | (0.202) |

**Figure 1C.**

**3D Search Motif #2: Non-heme Nitric Oxide:**

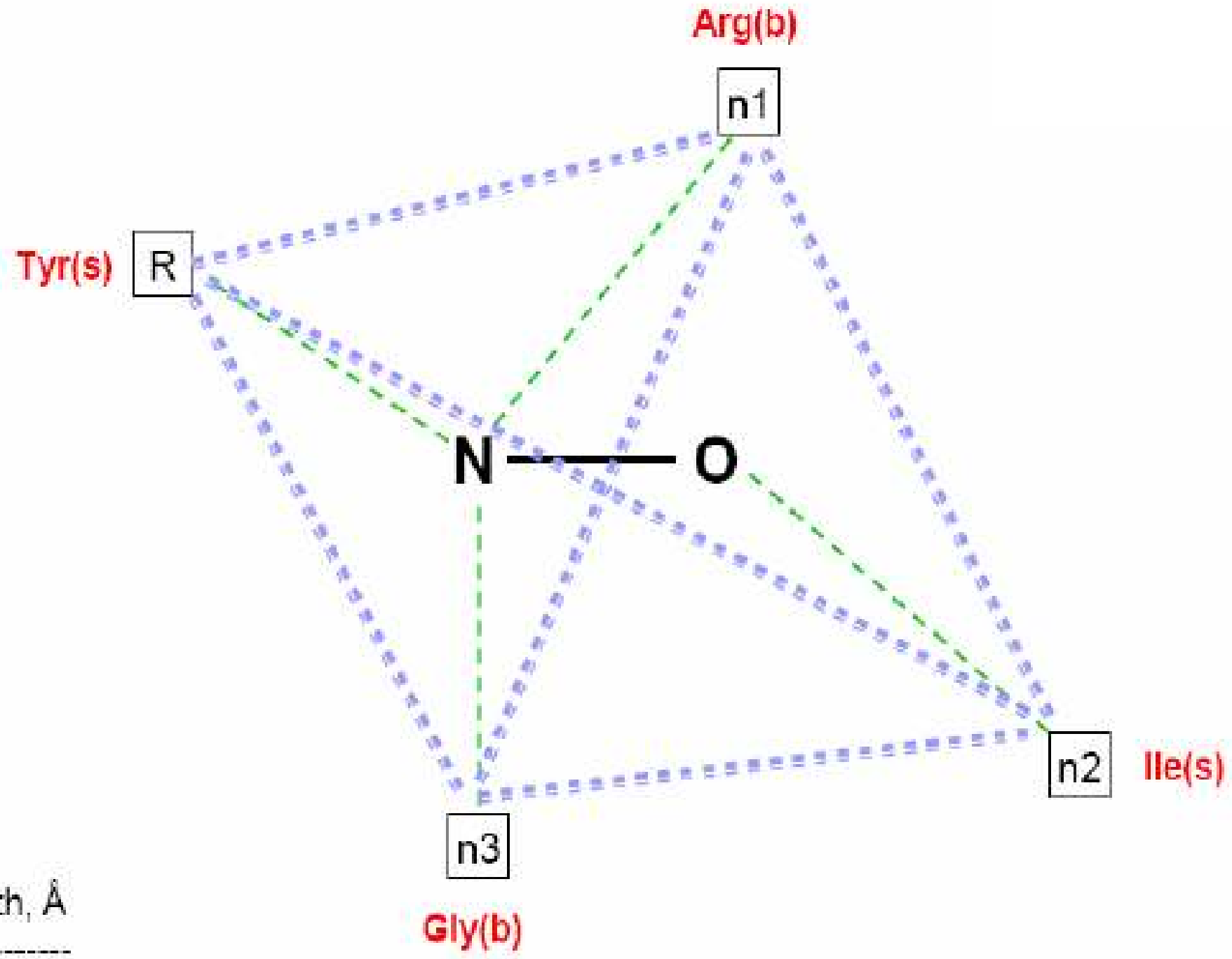

| side | length, Å | |
|---|---|---|
| Rn1 | 5.096 | (0.099) |
| Rn2 | 10.706 | (0.073) |
| Rn3 | 5.736 | (0.131) |
| n1n2 | 8.592 | (0.022) |
| n1n3 | 3.562 | (0.003) |
| n2n3 | 7.998 | (0.019) |

**Figure 1D.**

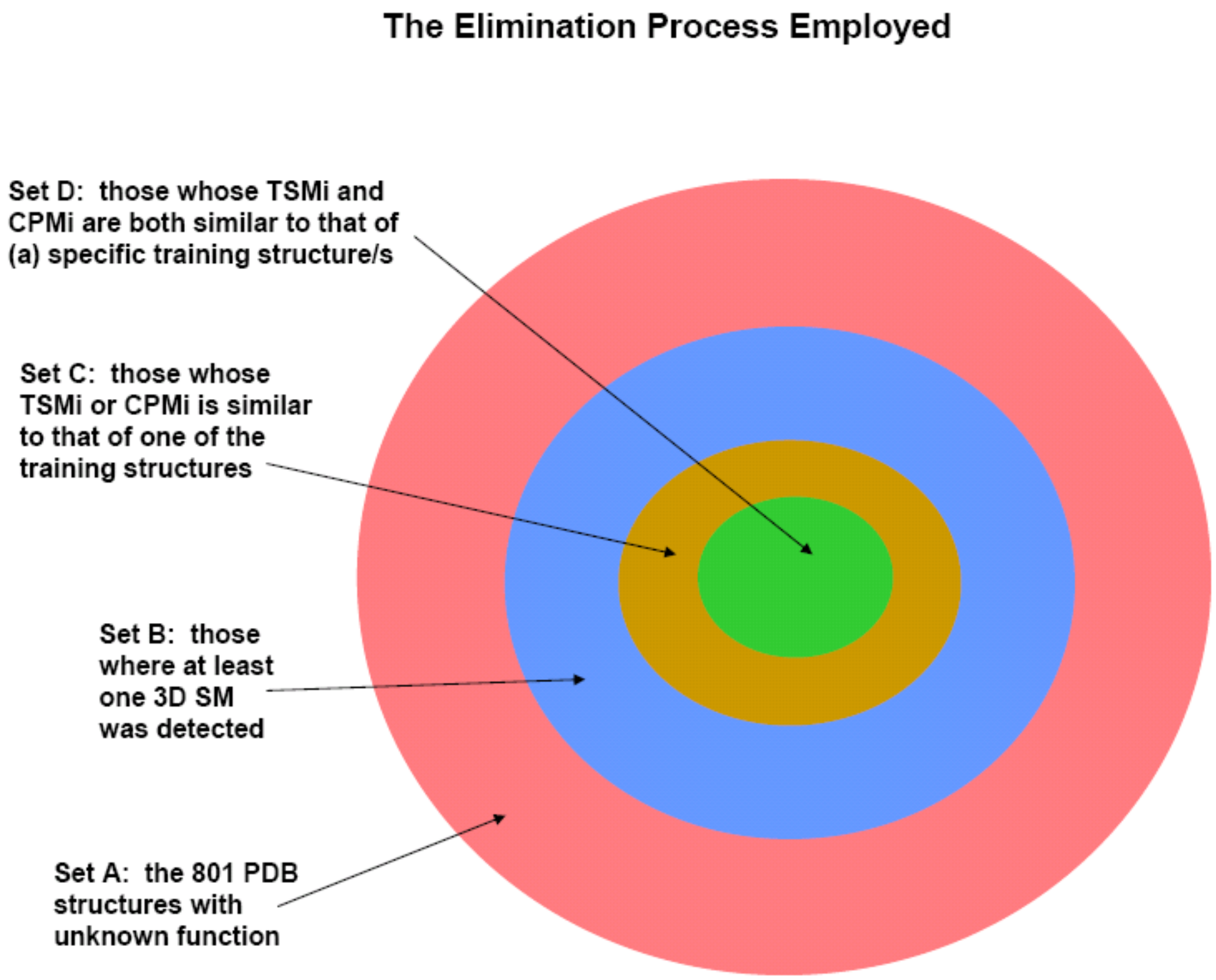
The Elimination Process Employed
Set D: those whose TSMi and CPMi are both similar to that of (a) specific training structure/s
Set C: those whose TSMi or CPMi is similar to that of one of the training structures
Set B: those where at least one 3D SM was detected
Set A: the 801 PDB structures with unknown function

**Figure 2.**

# TABLES:

## The Training Sets

| | | |
|---|---|---|
| **Sialic Acid Binding Site:** | | |
| 1JSN:A | *Influenza A virus* | Hemagglutinin HA1 chain (residues 1-325, chain A) and HA2 chain (residues 1-176, chain B) with bound N-acetyl-D-glucosamine (NAG), D-galactose (GAL), and O-sialic acid (SIA) |
| 1JSO:A | *Influenza A virus* | Hemagglutinin HA1 chain (residues 1-325; chain A) and HA2 chain (residues 1-176; chain B) with bound N-acetyl-D-glucosamine (NAG) and O-sialic acid (SIA) |
| 1W0O:A | *Vibrio cholerae* | Sialidase (E.C. 3.2.1.18; syn.: neuraminidase, nanase) with bound calcium ion, 2-deoxy-2,3-dehydro-N-acetyl-neuraminic acid (DAN) and O-sialic acid (SIA) |
| 1W0P:A | *Vibrio cholerae* | Sialidase (E.C. 3.2.1.18; syn.: neuraminidase, nanase) with bound calcium ion, glycerol (GOL), 2-amino-2-hydroxymethyl-propane-1,3-diol (TRS), and O-sialic acid (SIA) |
| 1MQN:A,D | *Influenza A virus* | Hemagglutinin HA1 chain (chains A, D, G) and HA2 chain (chains B, E, H) with bound N-acetyl-D-glucosamine (NAG), alpha-D-mannose (MAN), D-galactose (GAL) and O-sialic acid (SIA) molecules |
| **Retinoic Acid Binding Site:** | | |
| 1FM9:A | *Homo sapiens* | The heterodimer of the human RXR-alpha and PPAR-gamma ligand binding domains respectively bound with 9-cis retinoic acid and GI262570 and co-activator peptides |
| 1K74:A | *Homo sapiens* | The heterodimer of the human PPAR-gamma and RXR-alpha ligand binding domains respectively bound with GW409544 and 9-cis retinoic acid and co-activator peptides |
| 1FBY:A,B | *Homo sapiens* | The human RXR-alpha ligand binding domain bound to 9-cis retinoic acid |
| 1FM6:A,U | *Homo sapiens* | The heterodimer of the human RXR-alpha and PPAR-gamma ligand binding domains respectively bound with 9-cis retinoic acid and rosiglitazone and co-activator peptides |
| 1XDK:A,E | *Mus musculus* | The RAR-beta/RXR-alpha ligand binding domain heterodimer in complex with 9-cis retinoic acid and a fragment of the TRAP220 co-activator |
| 1XLS:A,B,C,D | *Homo sapiens / Mus musculus* | Heterodimer of the human RXR-alpha ligand binding domain and the mouse orphan nuclear receptor NR1I3 (syn.: constitutive androstane receptor, CAR) bound to TCPOBOP and 9-cis retinoic acid and a TIF2 peptide containg the 3rd LXXLL motifs |
| 2ACL:A,C,E,G | *Homo sapiens* | Heterodimer of the human retonic acid RXR-alpha and the mouse oxysterols receptor LXR-alpha (syn.: nuclear orphan receptor LXR-alpha) with bound 1-benzyl-3-(4-methoxyphenylamino)-4-phenylpyrrole-2,5-dione (L05) and retinoic acid |
| **Heme-Bound Nitric Oxide:** | | |
| 1OZW:A,B | *Homo sapiens* | The ferric, ferrous and ferrous-nitrogen oxide (heme-complexed) forms of the human heme oxygenase-1 (E.C.1.14.99.3) |
| 1XK3:A,B | *Homo sapiens* | Heme oxygenase-1 (HO-1) Arg183Glu mutant with bound heme-complexed nitrogen oxide (NO) |
| 1OZL:A,B | *Homo sapiens* | The ferric, ferrous and ferrous-nitrogen oxide (heme-complexed) forms of the Asp140Ala mutant of human heme oxygenase-1 (E.C.1.14.99.3) |
| **Unbound Nitric Oxide:** | | |
| 1ZGN:A,B | *Homo sapiens* | Glutathione-s-transferase pi (syn.: GST Class Pi) with bound dinitrosyl-diglutathionyl iron complex |

**Table 1.**

| | | | | | | | | | |
|---|---|---|---|---|---|---|---|---|---|
| 1DI6 | 1J9L | 1LJ7 | 1NNQ | 1O6A | 1QVV | 1RZ2 | 1SYR | 1TU9 | 1V70 |
| 1DI7 | 1JAL | 1LJO | 1NNW | 1O6D | 1QVW | 1RZ3 | 1T06 | 1TUA | 1V8D |
| 1DM5 | 1JN1 | 1LPL | 1NNX | 1O89 | 1QVZ | 1S12 | 1T07 | 1TUH | 1V8H |
| 1EW4 | 1JO0 | 1LQL | 1NO5 | 1O8C | 1QW2 | 1S2X | 1T0B | 1TUV | 1V8O |
| 1F89 | 1JOG | 1LXJ | 1NOG | 1ON0 | 1QY9 | 1S4C | 1T0T | 1TUW | 1V8P |
| 1FL9 | 1JOP | 1LXN | 1NPD | 1OQ1 | 1QYA | 1S4K | 1T1J | 1TWU | 1V96 |
| 1FUX | 1JOV | 1M1S | 1NPY | 1ORU | 1QYI | 1S5A | 1T2B | 1TWY | 1V99 |
| 1G2R | 1JRI | 1M33 | 1NQM | 1OSC | 1QZ4 | 1S7H | 1T3U | 1TXJ | 1V9B |
| 1H2H | 1JRK | 1M3S | 1NQN | 1OY1 | 1QZ8 | 1S7I | 1T57 | 1TXL | 1VAJ |
| 1HQQ | 1JSX | 1M65 | 1NR9 | 1OYZ | 1R0U | 1S7O | 1T5J | 1TXZ | 1VBK |
| 1HRU | 1JX7 | 1M68 | 1NRI | 1OZ9 | 1R3D | 1S8N | 1T5R | 1TY8 | 1VBV |
| 1HTW | 1JYH | 1M98 | 1NRK | 1P1L | 1R4V | 1S9U | 1T5Y | 1TZ0 | 1VCT |
| 1HXL | 1JZT | 1MK4 | 1NS5 | 1P1M | 1R5X | 1SAW | 1T62 | 1TZA | 1VDH |
| 1HXZ | 1K26 | 1ML8 | 1NU0 | 1P5F | 1R6Y | 1SBK | 1T6A | 1TZZ | 1VDW |
| 1HY2 | 1K2E | 1MOG | 1NX4 | 1P8C | 1R75 | 1SC0 | 1T6S | 1U05 | 1VE3 |
| 1I36 | 1K3R | 1MW5 | 1NX8 | 1P99 | 1R7L | 1SD5 | 1T6T | 1U0K | 1VGG |
| 1I60 | 1K4N | 1MW7 | 1NXH | 1P9I | 1RC6 | 1SDI | 1T8H | 1U5W | 1VGY |
| 1I6N | 1K77 | 1MWQ | 1NXJ | 1P9Q | 1RCU | 1SDJ | 1T95 | 1U61 | 1VH0 |
| 1I9H | 1K7J | 1MWW | 1NXZ | 1PB0 | 1RFE | 1SED | 1T9F | 1U69 | 1VH5 |
| 1IHN | 1K7K | 1MZG | 1NY1 | 1PBJ | 1RFZ | 1SEF | 1TC5 | 1U6L | 1VH6 |
| 1IJ8 | 1K8F | 1N1Q | 1NYE | 1PC6 | 1RI6 | 1SF9 | 1TD6 | 1U7I | 1VH9 |
| 1ILV | 1KJN | 1N81 | 1NZA | 1PD3 | 1RKI | 1SFN | 1TE5 | 1U7N | 1VHC |
| 1IN0 | 1KK9 | 1NC5 | 1NZJ | 1PF5 | 1RKQ | 1SFS | 1TEL | 1U84 | 1VHE |
| 1IUJ | 1KON | 1NC7 | 1NZN | 1PG6 | 1RLH | 1SFX | 1TLJ | 1U9C | 1VHF |
| 1IUK | 1KQ3 | 1NE2 | 1O0I | 1PM3 | 1RLJ | 1SG9 | 1TLQ | 1U9D | 1VHK |
| 1IUL | 1KQ4 | 1NE8 | 1O13 | 1PQY | 1RLK | 1SH8 | 1TO0 | 1U9P | 1VHM |
| 1IXL | 1KR4 | 1NF2 | 1O1Y | 1PT5 | 1RTT | 1SHE | 1TO3 | 1UAN | 1VHN |
| 1IZM | 1KUU | 1NG6 | 1O22 | 1PT7 | 1RTW | 1SJ5 | 1TOV | 1UC2 | 1VHO |
| 1J27 | 1KYH | 1NI9 | 1O3U | 1PT8 | 1RTY | 1SMB | 1TP6 | 1UCR | 1VHQ |
| 1J2R | 1KYT | 1NIG | 1O4T | 1PUG | 1RU8 | 1SPV | 1TPX | 1UE8 | 1VHS |
| 1J2V | 1L0B | 1NIJ | 1O4W | 1PV5 | 1RV9 | 1SQ4 | 1TQ5 | 1UF3 | 1VHU |
| 1J31 | 1L1S | 1NJH | 1O50 | 1PVM | 1RVK | 1SQE | 1TQ8 | 1UF9 | 1VHY |
| 1J3M | 1L5X | 1NJK | 1O51 | 1PW5 | 1RW0 | 1SQH | 1TQB | 1UFA | 1VI1 |
| 1J3W | 1L6R | 1NJR | 1O5J | 1Q2Y | 1RW1 | 1SQS | 1TQC | 1UFB | 1VI3 |
| 1J5U | 1LCV | 1NKQ | 1O5U | 1Q4R | 1RW7 | 1SQU | 1TQX | 1UFH | 1VI4 |
| 1J74 | 1LCW | 1NKV | 1O61 | 1Q77 | 1RXD | 1SQW | 1TSJ | 1UJ8 | 1VI7 |
| 1J7D | 1LCZ | 1NMN | 1O62 | 1Q7H | 1RXH | 1SR0 | 1TT4 | 1UMJ | 1VI8 |
| 1J8B | 1LDO | 1NMO | 1O65 | 1Q8B | 1RXJ | 1SS4 | 1TT7 | 1V30 | 1VIM |
| 1J9J | 1LDQ | 1NMP | 1O67 | 1Q8C | 1RXK | 1SU0 | 1TTZ | 1V6H | 1VIV |
| 1J9K | 1LEL | 1NNH | 1O69 | 1Q9U | 1RYL | 1SU1 | 1TU1 | 1V6T | 1VIZ |

**Table 2 (part 1 of 2).**

| | | | | | | | | | |
|---|---|---|---|---|---|---|---|---|---|
| 1VJ1 | 1VQW | 1WV8 | 1XKF | 1Y7I | 1YOA | 1ZC6 | 2A1V | 2ARZ | 2CUW |
| 1VJ2 | 1VQY | 1WV9 | 1XKL | 1Y7M | 1YOC | 1ZCE | 2A2L | 2ASF | 2CV9 |
| 1VJF | 1VQZ | 1WVI | 1XKQ | 1Y7P | 1YOX | 1ZD0 | 2A2M | 2ATR | 2CVB |
| 1VJG | 1VR4 | 1WWI | 1XM5 | 1Y7R | 1YOY | 1ZE0 | 2A2O | 2ATZ | 2CVE |
| 1VJK | 1VR9 | 1WWP | 1XM7 | 1Y80 | 1YOZ | 1ZEE | 2A33 | 2AU5 | 2CVL |
| 1VJL | 1VRM | 1WWZ | 1XMT | 1Y81 | 1YQE | 1ZHV | 2A35 | 2AUA | 2CW4 |
| 1VJU | 1W8I | 1WY6 | 1XMX | 1Y82 | 1YQF | 1ZKD | 2A3N | 2AUW | 2CW5 |
| 1VJX | 1W9A | 1X6I | 1XN4 | 1Y88 | 1YQH | 1ZKE | 2A3Q | 2AV4 | 2CWQ |
| 1VK0 | 1WD5 | 1X6J | 1XPJ | 1Y89 | 1YRE | 1ZKI | 2A5Z | 2AVN | 2CWY |
| 1VK1 | 1WD6 | 1X72 | 1XQ4 | 1Y8A | 1YS9 | 1ZKP | 2A67 | 2AX3 | 2CX0 |
| 1VK5 | 1WDI | 1X77 | 1XQ6 | 1Y8T | 1YTL | 1ZL0 | 2A6B | 2AXO | 2CX1 |
| 1VK8 | 1WDJ | 1X7F | 1XQ9 | 1Y9B | 1YUD | 1ZMB | 2A6C | 2AXP | 2D2Y |
| 1VK9 | 1WDT | 1X7V | 1XQA | 1Y9E | 1YV9 | 1ZN6 | 2A8E | 2AZ4 | 2D4R |
| 1VKA | 1WDV | 1X9G | 1XQB | 1Y9I | 1YW1 | 1ZNP | 2A9F | 2AZP | 2ES9 |
| 1VKB | 1WEH | 1XA0 | 1XRG | 1YAC | 1YW3 | 1ZOX | 2A9S | 2B0A | 2ESH |
| 1VKD | 1WEK | 1XAF | 1XRI | 1YAV | 1YWF | 1ZP6 | 2AAM | 2B0C | 2ESN |
| 1VKH | 1WHZ | 1XB4 | 1XSV | 1YB2 | 1YX1 | 1ZPV | 2AB0 | 2B0R | 2ETD |
| 1VKI | 1WJ9 | 1XBF | 1XTL | 1YB3 | 1YYV | 1ZPW | 2AB1 | 2B0V | 2ETH |
| 1VKM | 1WK2 | 1XBV | 1XTM | 1YBM | 1YZV | 1ZPY | 2ACA | 2B1Y | 2ETS |
| 1VKW | 1WK4 | 1XBW | 1XTO | 1YBX | 1YZY | 1ZQ7 | 2AEG | 2B2P | 2EUC |
| 1VL0 | 1WKC | 1XBX | 1XUV | 1YCD | 1YZZ | 1ZS7 | 2AEU | 2B2Z | 2EUI |
| 1VL4 | 1WLU | 1XBY | 1XV2 | 1YCY | 1Z0P | 1ZSO | 2AEV | 2B30 | 2EVE |
| 1VL5 | 1WLV | 1XBZ | 1XVS | 1YDF | 1Z1S | 1ZSW | 2AFC | 2B33 | 2EVR |
| 1VL7 | 1WLZ | 1XCC | 1XW8 | 1YDH | 1Z40 | 1ZTC | 2AH5 | 2B3M | 2EVV |
| 1VLY | 1WM6 | 1XDI | 1XWM | 1YDM | 1Z67 | 1ZTD | 2AH6 | 2B3N | 2EW0 |
| 1VM0 | 1WMM | 1XE1 | 1XX7 | 1YDW | 1Z6M | 1ZTP | 2AI4 | 2B41 | 2EWC |
| 1VMF | 1WN3 | 1XE7 | 1XXL | 1YE5 | 1Z6N | 1ZTV | 2AJ2 | 2B4A | 2EWR |
| 1VMH | 1WN9 | 1XE8 | 1XY7 | 1YEM | 1Z7A | 1ZUP | 2AJ6 | 2B4W | 2F06 |
| 1VMJ | 1WNA | 1XFI | 1Y0H | 1YEY | 1Z7U | 1ZVP | 2AJ7 | 2B6C | 2F20 |
| 1VP2 | 1WOL | 1XFJ | 1Y0K | 1YF9 | 1Z84 | 1ZWJ | 2ALI | 2B6E | 2F22 |
| 1VP4 | 1WOZ | 1XFS | 1Y0N | 1YHF | 1Z85 | 1ZWY | 2AMH | 2B8M | 2F4L |
| 1VP8 | 1WPB | 1XG7 | 1Y0Z | 1YKW | 1Z8H | 1ZX3 | 2AMU | 2BBE | 2F4N |
| 1VPB | 1WR2 | 1XG8 | 1Y12 | 1YLK | 1Z90 | 1ZX5 | 2AO9 | 2BDT | 2F4Z |
| 1VPH | 1WSC | 1XHN | 1Y1X | 1YLL | 1Z94 | 1ZX8 | 2AP3 | 2BDV | 2F9C |
| 1VPQ | 1WTY | 1XHO | 1Y2I | 1YLM | 1Z9T | 1ZXJ | 2AP6 | 2BE4 | 2FBL |
| 1VPV | 1WU8 | 1XI6 | 1Y5H | 1YLN | 1ZBM | 1ZXO | 2APJ | 2C5Q | 2FBM |
| 1VPY | 1WUE | 1XI8 | 1Y63 | 1YLO | 1ZBO | 1ZXU | 2APL | 2COH | 2FDS |
| 1VPZ | 1WUF | 1XIZ | 1Y6Z | 1YLX | 1ZBP | 1ZZM | 2AQW | 2CSL | 2FE1 |
| 1VQR | 1WUS | 1XJC | 1Y71 | 1YN4 | 1ZBR | 2A13 | 2AR1 | 2CU5 | 2FFG |
| 1VQS | 1WV3 | 1XK8 | 1Y7H | 1YN5 | 1ZBS | 2A15 | 2ARH | 2CU6 | 2FFI |
| | | | | | | | | | 2FFM |

**Table 2  (part 2 of 2).**

| PDB ID* | DESCRIPTION |
|---|---|
| 104M | Sperm whale (*Physeter catodon* ) skeletal muscle myoglobin (heme-iron[II]-bound) with bound N-butyl isocyanide and sulfate ion at pH 7.0 |
| 1ASH | Iron(II)-protoporphyrin IX-bound hemoglobin domain I from *Ascaris suum* with bound dioxygen at 2.2 Å resolution |
| 1B3B | Structure of glutamate dehydrogenase from *Thermotoga maritima* with mutations N97D and G376K |
| 1BRF | Structure of Rubredoxin with bound Fe(III) from Pyrococcus furiosus at 0.95 Å resolution |
| 1CBN | Structure of the hydrophobic protein crambin from the seed of *Crambe abyssinica* (Abyssinian cabbage) at 130°K and at 0.83 Å resolution |
| 1CKO | Structure of mRNA capping enzyme from *Chlorella* virus PBCV-1 in complex with the CAP analog GpppG |
| 1CRP | NMR structure (n=20) of human C-H-Ras p21 protein (catalytic domain, res. 1-166) complexed with GDP and MG |
| 1EWK | Structure of the metabotropic glutamate receptor subtype 1 from *Rattus norvegicus* complexed with glutamate |
| 1F3O | Structure of MJ0796 ATP-binding cassette with bound Mg-ADP from *Methanococcus jannaschii* |
| 1FW5 | Solution structure of membrane binding peptide of Semliki forest virus mRNA capping enzyme NSP1 |
| 1HWY | Glutamate dehydrogenase from *Bos taurus* complexed with NAD and 2-oxoglutarate |
| 1JFF | Refined structure of bovine (Bos taurus) α-β tubulin from zinc-induced sheets stabilized with taxol |
| 1MJJ | Structure of the complex of the Fab fragment of esterolytic antibody MS6-12 and the transition-state analog, N-{[2-({[1-(4-carboxybutanoyl)amino]-2-phenylethyl}-hydroxyphosphinyl)oxy]acetyl}-2-phenylethylamine |
| 1MV5 | Structure of the ATP-binding domain of the multidrug resistance ABC transporter and permease protein from *Lactococcus lactis* with bound ADP, ATP and Mg ion |
| 1NQT | Structure of glutamate dehydrogenase from *Bos taurus* with bound ADP |
| 1PE6 | Structure of papain (E.C.4.3.22.2) from the papaya fruit (*Carica papaya* ) latex complexed with E-64-C ((2S,3S)-3-(1-(N-(3-methylbutyl)amino)-leucylcarboxyl)oxirane-2-carboxylate) at 2.1 Å resolution |
| 1RQ7 | *Mycobacterium tuberculosis* FTSZ (filamenting temperature-sensitive mutant Z) in complex with GDP |
| 1SVS | Structure of the K180P mutant of GI α subunit bound to GPPNHP (phosphoaminophosphonic acid-guanylate ester) |
| 1TUB | Electron diffraction structrue of *Sus scrofa* (pig) tubulin α-β dimer with bound GTP, GDP and taxotere |
| 1TWY | Structure of a hypothetical ABC-type phosphate transporter from *Vibrio cholerae* O1 Biovar eltor |
| 1Z3C | mRNA cap (guanine-N7) methyltransferasein from the encephalitozooan *Cuniculi* complexed with AzoAdoMet |

*These negative control structres were used for sialic acid, retinoic acid, and heme-bound and unbound nitric oxide; negative control structures used for ATP (ser/thr protein kinases) and GTP (small, Ras-type G-proteins) are enumerated and discussed in Reyes, V.M., 2008a.

**Table 3.**

## Species Distribution of the 801 Application Structures

| Source organism | No. structures | % of Total | Source organism | No. structures | % of Total |
|---|---|---|---|---|---|
| Escherichia coli | 88 | 10.99 | Xanthomonas campestris | 02 | 0.25 |
| Thermotoga maritima | 63 | 7.86 | Vibrio parahaemolyticus | 02 | 0.25 |
| Thermus thermophilus | 56 | 6.99 | Sulfolobus solfataricus | 02 | 0.25 |
| Bacillus subtilis | 48 | 5.99 | Streptococcus mutans | 02 | 0.25 |
| Pseudomonas aeruginosa | 38 | 4.74 | Salmonella enterica | 02 | 0.25 |
| Haemophilus influenzae | 26 | 3.24 | Rattus norvegicus | 02 | 0.25 |
| Archaeoglobus fulgidus | 23 | 2.87 | Pseudomonas putida | 02 | 0.25 |
| Pyrococcus furiosus | 20 | 2.50 | Plasmodium yoelii | 02 | 0.25 |
| Pyrococcus horikoshii | 20 | 2.50 | Neisseria meningitidis | 02 | 0.25 |
| Arabidopsis thaliana | 20 | 2.50 | Methanosarcina mazei | 02 | 0.25 |
| Staphylococcus aureus | 18 | 2.25 | Erwinia carotovora | 02 | 0.25 |
| Homo sapiens | 18 | 2.25 | Deinococcus radiodurans | 02 | 0.25 |
| Saccharomyces cerevisiae | 17 | 2.12 | Chromobacterium violaceum | 02 | 0.25 |
| Mycobacterium tuberculosis | 16 | 1.99 | Caulobacter crescentus | 02 | 0.25 |
| Vibrio cholerae | 13 | 1.62 | Capra hircus | 02 | 0.25 |
| Enterococcus faecalis | 13 | 1.62 | Bordetella bronchiseptica | 02 | 0.25 |
| Bacillus cereus | 13 | 1.62 | Bacillus anthracis | 02 | 0.25 |
| Agrobacterium tumefaciens | 13 | 1.62 | Uncultured bacterium | 01 | 0.12 |
| Streptomyces avidinii | 12 | 1.50 | Trypanosoma brucei | 01 | 0.12 |
| Thermoplasma acidophilum | 11 | 1.37 | Trypanosoma cruzi | 01 | 0.12 |
| Leishmania major | 10 | 1.25 | Toxoplasma gondii | 01 | 0.12 |
| Bacillus stearothermophilus | 10 | 1.25 | Synthetic protein | 01 | 0.12 |
| Streptococcus pyogenes | 08 | 1.00 | Streptomyces glaucescens | 01 | 0.12 |
| Streptococcus pneumoniae | 08 | 1.00 | Schizosaccharomyces pombe | 01 | 0.12 |
| Salmonella typhimurium | 08 | 1.00 | Sars coronavirus | 01 | 0.12 |
| Pyrobaculum aerophilum | 08 | 1.00 | Rhodopseudomonas palustris | 01 | 0.12 |
| Plasmodium falciparum | 07 | 0.87 | Pseudomonas syringae | 01 | 0.12 |
| Nitrosomonas europaea | 07 | 0.87 | Plasmodium berghei | 01 | 0.12 |
| Methanobacterium thermoautotrophicum | 07 | 0.87 | Plasmodium vivax | 01 | 0.12 |
| Bacteroides thetaiotaomicron | 07 | 0.87 | Plasmodium yoelli | 01 | 0.12 |
| Bacillus halodurans | 07 | 0.87 | Plasmodium knowlesi | 01 | 0.12 |
| Shewanella oneidensis | 06 | 0.75 | Mycoplasma genitalium | 01 | 0.12 |
| Methanococcus jannaschii | 06 | 0.75 | Moorella thermoacetica | 01 | 0.12 |
| Helicobacter pylori | 06 | 0.75 | Listeria monocytogenes | 01 | 0.12 |
| Aquifex aeolicus | 06 | 0.75 | Listeria innocua | 01 | 0.12 |
| Sulfolobus tokodaii | 05 | 0.62 | Leishmania donovani | 01 | 0.12 |
| Shigella flexneri | 05 | 0.62 | Klebsiella pneumoniae | 01 | 0.12 |
| Ovis aries | 05 | 0.62 | Influenza A | 01 | 0.12 |
| Mus musculus | 05 | 0.62 | Hydra vulgaris | 01 | 0.12 |
| Methanothermobacter | 05 | 0.62 | Halobacterium salinarum | 01 | 0.12 |
| Gallus gallus | 05 | 0.62 | Drosophila melanogaster | 01 | 0.12 |
| Caenorhabditis elegans | 05 | 0.62 | Desulfovibrio vulgaris | 01 | 0.12 |
| Aeropyrum pernix | 05 | 0.62 | Danio rerio | 01 | 0.12 |
| Nostoc punctiforme | 04 | 0.50 | Cryptosporidium parvum | 01 | 0.12 |
| Clostridium acetobutylicum | 04 | 0.50 | Citrobacter braakii | 01 | 0.12 |
| Campylobacter jejuni | 04 | 0.50 | Camelus dromedarius | 01 | 0.12 |
| Bacillus stearothermophilus | 04 | 0.50 | Bradyrhizobium japonicum | 01 | 0.12 |
| Porphyromonas gingivalis | 03 | 0.37 | Bordetella parapertussis | 01 | 0.12 |
| Nicotiana tabacum | 03 | 0.37 | Bordetella pertussis | 01 | 0.12 |
| Mycoplasma pneumoniae | 03 | 0.37 | Bacteriophage lambda | 01 | 0.12 |
| Escherichia coli & Shigella | 03 | 0.37 | Bacillus brevis | 01 | 0.12 |
| Clostridium thermocellum | 03 | 0.37 | Arthrospira maxima | 01 | 0.12 |
| Chlorobium tepidum | 03 | 0.37 | Acinetobacter sp. | 01 | 0.12 |

**Table 4.**

## GTP-binding small Ras-type G-proteins

| CPM | | | TSM | | |
|---|---|---|---|---|---|
| | 1o3y:A | 8.0330 | | 1m7b:A | 16.6906 |
| | 1nvu:Q | 10.2041 | | 1e96:A | 17.7361 |
| | 1n61:A | 10.3577 | | 1loo:A | 23.0242 |
| | 2rap:_ | 10.6095 | | 1n61:A | 27.6453 |
| | 1m7b:A | 16.4748 | | 1nvu:Q | 30.6122 |
| | 1e96:A | 16.5826 | | 2rap:_ | 31.3017 |
| | 1loo:A | 16.7015 | | 1o3y:A | 31.6817 |
| | 1vhq:B | 2.6710 | | 2b30:C | 2.4272 |
| * | 1vhq:A | 3.0928 | | 2b30:D | 2.4713 |
| * | 1oy1:D | 3.2967 | | 2b30:B | 2.5652 |
| * | 1vim:C | 3.3058 | | 2b30:A | 2.6979 |
| * | 1oy1:C | 3.4216 | * | 1xtl:B | 18.3979 |
| * | 1oy1:B | 4.3478 | * | 1ru8:A | 19.4886 |
| * | 1sg9:B | 5.6351 | * | 1ru8:B | 19.4886 |
| * | 1oy1:A | 5.6747 | | 1vim:C | 42.4931 |
| * | 1to0:B | 6.0124 | | 1sg9:B | 45.7737 |
| * | 1to0:G | 6.3660 | | 1oy1:A | 51.0719 |
| * | 1ru8:A | 7.5000 | | 1oy1:B | 53.7084 |
| * | 1ru8:B | 7.5000 | | 1oy1:D | 56.2379 |
| * | 1to0:D | 8.5863 | | 1oy1:C | 56.6172 |
| * | 1xtl:B | 17.7817 | | 1vhq:A | 56.8299 |
| | 2b30:A | 25.6966 | | 1vhq:B | 58.3713 |
| | 2b30:B | 25.9620 | | 1to0:D | 66.5221 |
| | 2b30:C | 26.3019 | | 1to0:G | 75.5084 |
| | 2b30:D | 26.3019 | | 1to0:B | 76.7462 |

| | | | | | | |
|---|---|---|---|---|---|---|
| | 1oy1 | A-219 | K-132 | G-131 | D-90 | AAAA |
| | 1oy1 | A-219 | K-132 | G-131 | D-90 | BBBB |
| | 1oy1 | A-219 | K-132 | G-131 | D-90 | CCCC |
| | 1oy1 | A-219 | K-132 | G-131 | D-90 | DDDD |
| * | 1ru8 | A-59 | K-68 | G-69 | D-56 | AAAA |
| * | 1ru8 | A-59 | K-68 | G-69 | D-56 | BBBB |
| | 1sg9 | G-248 | K-202 | G-225 | D-250 | BBBB |
| | 1to0 | G-60 | K-56 | G-112 | D-57 | BBBB |
| | 1to0 | G-60 | K-56 | G-112 | D-57 | DDDD |
| | 1to0 | G-60 | K-56 | G-112 | D-57 | GGGG |
| | 1vhq | A-219 | K-132 | G-131 | D-90 | AAAA |
| | 1vim | A-118 | K-112 | G-87 | D-114 | CCCC |
| * | 1xtl | A-169 | K-53 | G-52 | D-171 | BBBB |

**Table 5 (part 1 of 7)**

**ATP-binding sert/thr Protein Kinases**

**CPM**

| | | | | | |
|---|---|---|---|---|---|
| 1fin:A | -0.0000 | 2phk:A | 16.9862 | 1gol:_ | 17.9938 |
| 1fin:C | -0.0000 | 1phk:_ | 17.5864 | 1ql6:A | 18.3260 |
| 1qmz:A | -0.0000 | 1hck:_ | 17.5949 | 1jst:C | 19.3643 |
| 1qmz:C | -0.0000 | 1b39:A | 17.6445 | 1jst:A | 19.6570 |
| | | 1b39:A | 17.8587 | | |

[illegible]

**TSM**

| | | | |
|---|---|---|---|
| 1gol:_ | 17.5801 | 1jst:C | 23.0029 |
| 1ql6:A | 17.8855 | 1jst:A | 23.8812 |
| 2phk:A | 19.3045 | 1fin:A | -100.0000 |
| 1phk:_ | 19.5603 | 1fin:C | -100.0000 |
| 1hck:_ | 21.3080 | 1qmz:A | -100.0000 |
| 1b39:A | 22.3555 | 1qmz:C | -100.0000 |
| 1b38:A | 22.6552 | | |

[illegible]

[NOTE: only structures w/ CPM and TSM values within 10.0 units of those of training structures are shown]

| | | | | | |
|---|---|---|---|---|---|
| 1f19 | V-71 | E-94 | L-86 | E-113 | AAAA |
| 1lxn | V-34 | E-55 | L-50 | E-48 | BBBB |
| 1mww | V-97 | D-58 | L-90 | E-85 | AAAA |
| 1mww | V-97 | D-58 | L-90 | E-85 | BBBB |
| 1mww | V-97 | D-58 | L-90 | E-85 | CCCC |
| 1nf2 | V-51 | E-19 | L-13 | D-10 | AAAA |
| 1nf2 | V-351 | E-319 | L-313 | D-310 | BBBB |
| 1nf2 | V-651 | E-619 | L-613 | D-610 | CCCC |
| 1o1y | V-134 | E-175 | L-213 | E-209 | AAAA |
| 1pm3 | V-26 | E-5 | M-1 | D-29 | BBBB |
| 1r5x | V-33 | E-13 | L-7 | E-36 | AAAA |
| 1r5x | V-33 | E-13 | L-7 | E-36 | BBBB |
| 1rki | V-44 | D-29 | L-48 | E-52 | AAAA |
| 1rkq | V-238 | D-259 | L-16 | D-12 | AAAA |
| 1rkq | V-238 | D-259 | L-16 | D-12 | BBBB |
| 1sbk | V-59 | E-84 | L-131 | E-84 | DDDD |
| 1t57 | V-172 | E-117 | L-95 | E-131 | CCCC |
| 1tq8 | V-124 | D-152 | L-154 | D-152 | AAAA |
| 1tq8 | V-124 | D-152 | L-154 | D-152 | BBBB |
| 1tq8 | V-124 | D-152 | L-154 | D-152 | CCCC |
| 1tq8 | V-124 | D-152 | L-154 | D-152 | DDDD |
| 1tq8 | V-124 | D-152 | L-154 | D-152 | FFFF |
| 1tt7 | V-38 | D-68 | L-65 | D-99 | BBBB (1) |
| 1tt7 | V-38 | D-68 | L-65 | D-99 | BBBB (2) |
| 1tt7 | V-96 | D-68 | L-65 | D-99 | BBBB (3) |
| 1tt7 | V-96 | D-68 | L-65 | D-99 | BBBB (4) |
| 1tt7 | V-38 | D-68 | L-65 | D-99 | CCCC (1) |
| 1tt7 | V-96 | D-68 | L-65 | D-99 | CCCC (2) |
| 1u9p | V-52 | E-59 | L-77 | E-73 | AAAA |
| 1ucr | V-9 | E-2 | L-46 | E-50 | AAAA (2) |
| 1ucr | V-9 | E-53 | L-46 | E-50 | AAAA (3) |
| 1ucr | V-9 | E-2 | L-46 | E-50 | BBBB (2) |
| 1ucr | V-9 | E-53 | L-46 | E-50 | BBBB (3) |
| 1vk0 | V-109 | E-157 | L-43 | E-39 | AAAA |
| 1vk0 | V-109 | E-157 | L-43 | E-39 | CCCC |
| 1vk0 | V-109 | E-157 | L-43 | E-39 | DDDD |
| 1vph | V-100 | E-133 | L-135 | E-133 | CCCC |
| 1vph | V-100 | E-133 | L-135 | E-133 | FFFF |
| 1wm6 | V-13 | E-81 | M-1 | D-3 | AAAA |
| 1wm6 | V-13 | D-3 | L-16 | E-20 | FFFF (1) |
| 1wm6 | V-13 | D-3 | L-53 | E-20 | FFFF (2) |
| 1wm6 | V-23 | D-3 | L-16 | E-20 | FFFF (3) |
| 1xbv | V-47 | D-62 | L-38 | E-42 | AAAA |
| 1xbv | V-47 | D-62 | L-38 | E-42 | BBBB |
| 1xbx | V-47 | D-62 | L-38 | E-42 | AAAA |
| 1xbx | V-47 | D-62 | L-38 | E-42 | BBBB |
| 1xby | V-47 | D-62 | L-38 | E-42 | AAAA |
| 1xbz | V-47 | D-62 | L-38 | E-42 | AAAA |
| 1xrg | V-78 | E-119 | L-117 | D-82 | AAAA (1) |
| 1y9e | V-38 | D-68 | L-65 | D-99 | AAAA (1) |
| 1y9e | V-38 | D-68 | L-65 | D-99 | AAAA (2) |
| 1y9e | V-38 | D-68 | L-65 | D-99 | BBBB (1) |
| 1y9e | V-38 | D-68 | L-65 | D-99 | BBBB (2) |
| 1y9e | V-38 | D-68 | L-65 | D-99 | DDDD |
| 1yyv | V-99 | E-97 | L-76 | D-73 | AAAA |
| 1yyv | V-99 | E-97 | L-76 | D-73 | BBBB |
| 2c5q | V-229 | D-225 | L-208 | E-204 | AAAA (1) |
| 2c5q | V-229 | D-225 | L-208 | E-204 | AAAA (2) |
| 2c5q | V-229 | D-225 | L-208 | E-204 | BBBB |
| 2c5q | V-229 | D-225 | L-208 | E-204 | CCCC |
| 2c5q | V-229 | D-225 | L-208 | E-204 | DDDD |
| 2c5q | V-229 | D-225 | L-208 | E-204 | EEEE |
| 2cvl | V-104 | E-116 | L-77 | E-116 | BBBB |
| 2cvl | V-104 | E-116 | L-77 | E-116 | CCCC |
| 2cvl | V-104 | E-116 | L-77 | E-116 | FFFF |
| 2f41 | V-194 | E-32 | L-34 | D-173 | AAAA |
| 2f41 | V-194 | E-32 | L-34 | D-173 | BBBB |
| 2f41 | V-194 | E-32 | L-34 | D-173 | DDDD |

**Table 5 (part 2 of 7)**

## sialic acid-binding proteins

| CPM | | | | TSM | | | |
|---|---|---|---|---|---|---|---|
| 1w0o:A | 2.3854 | | | 1jso:A | 52.4113 | | |
| 1w0p:A | 2.3854 | | | 1jsn:A | 52.5309 | | |
| 1mqn:A | 14.1443 | | | 1mqn:A | 52.8660 | | |
| 1mqn:D | 14.3092 | | | 1mqn:D | 52.8783 | | |
| 1jsn:A | 15.1853 | | | 1w0o:A | 92.1400 | | |
| 1jso:A | 15.3448 | | | 1w0p:A | 92.2602 | | |
| * 1vka:A | 4.4976 | * 1t62:B | 16.4414 | 1y7p:A | 1.2232 | 1nxj:B | 18.3348 |
| * 1iuk:A | 5.1423 | * 1j7d:A | 16.4568 | 1y7p:C | 1.3810 | 1t62:B | 20.5706 |
| * 1sqh:A | 5.5626 | * 1j74:A | 16.6065 | 1pt8:B | 1.7413 | 1te5:B | 20.7348 |
| * 1f89:A | 7.2532 | * 1vdh:E | 16.7980 | 1pt8:A | 1.7680 | 1sfs:A | 23.2239 |
| * 1f89:B | 7.2532 | * 1qyi:A | 19.4749 | 1pt5:A | 1.7724 | 1t62:A | 25.3779 |
| * 1mzg:B | 7.7333 | * 1wue:B | 20.4061 | 1pt7:A | 1.7724 | 1uc2:B | 27.4541 |
| * 1ywf:A | 7.8041 | * 1wue:A | 20.6489 | 1pt7:B | 1.7724 | 1uc2:A | 27.8532 |
| * 1mzg:A | 8.4137 | * 1z94:F | 21.6438 | 1pt5:B | 1.8035 | 1f89:B | 28.6851 |
| * 1y6z:B | 8.6742 | * 1tc5:D | 24.7067 | 1tc5:A | 2.3320 | 1f89:A | 28.8255 |
| * 1nxj:B | 9.9203 | * 1tc5:C | 24.9828 | 1tc5:B | 2.3940 | 1j74:A | 31.5884 |
| * 1o62:A | 10.0508 | * 1tc5:B | 25.1026 | 1tc5:C | 2.5465 | 1mzg:A | 31.8142 |
| * 2ar1:A | 10.7170 | * 1tc5:A | 25.1029 | 1tc5:D | 2.6225 | 1j7d:A | 32.0144 |
| * 1te5:B | 11.5752 | 1wr2:A | 25.7388 | 1z94:F | 4.8402 | 1ywf:A | 32.9925 |
| * 1uc2:A | 12.1043 | 1y7p:C | 30.5085 | 1qyi:A | 7.7767 | 1vdh:E | 35.6158 |
| * 1uc2:B | 12.1841 | 1y7p:A | 31.4373 | 1wr2:A | 10.2633 | 1o62:A | 37.2927 |
| * 1t62:A | 12.6492 | 1pt7:B | 35.9453 | 1wk4:C | 14.4000 | 1vdh:D | 38.1281 |
| * 1yv9:A | 13.6226 | 1pt7:A | 36.1629 | 1wue:A | 14.4981 | 1mzg:B | 38.3111 |
| * 1wk4:C | 14.9818 | 1pt8:A | 36.6625 | 1wue:B | 14.5516 | 2ar1:A | 39.2444 |
| * 1sfs:A | 15.5316 | 1pt5:A | 36.7537 | 1yv9:A | 14.9849 | * 1vka:A | 40.0957 |
| * 1vdh:D | 15.5665 | 1pt5:B | 36.7848 | 1wuf:A | 17.2437 | * 1iuk:A | 53.0762 |
| * 1wuf:A | 16.0896 | 1pt8:B | 36.9403 | 1wuf:B | 17.2777 | * 1y6z:B | 62.9867 |
| * 1wuf:B | 16.1575 | | | | | * 1sqh:A | 64.3911 |

## sialic acid-binding proteins (cont'd.)

| | | | | | | |
|---|---|---|---|---|---|---|
| | 1f89 | R-226 | Y-235 | W-204 | G-236 | AAAA |
| | 1f89 | R-526 | Y-535 | W-504 | G-536 | BBBB |
| * | 1iuk | R-135 | Y-35 | M-127 | V-128 | AAAA |
| | 1j74 | R-61 | Y-63 | M-126 | G-52 | AAAA |
| | 1j7d | R-61 | Y-63 | M-126 | G-52 | AAAA |
| | 1mzg | Q-49 | S-47 | W-56 | V-55 | AAAA |
| | 1mzg | Q-49 | S-47 | W-56 | V-55 | BBBB |
| | 1nxj | R-130 | S-40 | W-90 | V-54 | BBBB |
| | 1o62 | Q-39 | S-40 | M-242 | V-41 | AAAA |
| | 1qyi | R-17 | Y-207 | W-184 | V-21 | AAAA |
| | 1sfs | Q-12 | S-10 | W-32 | V-14 | AAAA |
| * | 1sqh | Q-99 | S-97 | W-98 | G-80 | AAAA |
| | 1t62 | Q-1080 | S-1079 | M-1027 | G-1077 | AAAA |
| | 1t62 | Q-2080 | S-2079 | M-2027 | G-2077 | BBBB |
| | 1tc5 | Q-99 | Y-41 | M-39 | V-40 | AAAA |
| | 1tc5 | Q-99 | Y-41 | M-39 | V-40 | BBBB |
| | 1tc5 | Q-99 | Y-41 | M-39 | V-40 | CCCC |
| | 1tc5 | Q-99 | Y-41 | M-39 | V-40 | DDDD |
| | 1te5 | Q-239 | Y-191 | W-247 | V-245 | BBBB |
| | 1uc2 | R-27 | Y-29 | M-63 | V-62 | AAAA |
| | 1uc2 | R-27 | Y-29 | M-63 | V-62 | BBBB |
| | 1vdh | R-52 | Y-62 | W-51 | V-48 | DDDD |
| | 1vdh | R-52 | Y-62 | W-51 | V-48 | EEEE |
| * | 1vka | R-84 | S-55 | M-48 | G-81 | AAAA |
| | 1wk4 | R-80 | S-82 | W-128 | V-127 | CCCC |
| | 1wue | R-1235 | S-1259 | W-1288 | V-1287 | AAAA |
| | 1wue | R-2235 | S-2259 | W-2288 | V-2287 | BBBB |
| | 1wuf | Q-1088 | S-1271 | M-1270 | G-1269 | AAAA |
| | 1wuf | Q-2088 | S-2271 | M-2270 | G-2269 | BBBB |
| * | 1y6z | R-72 | Y-74 | W-5 | V-99 | BBBB |
| | 1yv9 | Q-219 | S-216 | M-189 | G-217 | AAAA |
| | 1ywf | R-235 | Y-238 | M-126 | G-230 | AAAA |
| | 1z94 | Q-136 | S-133 | W-130 | G-129 | FFFF |
| | 2ar1 | R-77 | Y-80 | M-106 | V-107 | AAAA |

**Table 5 (part 3 of 7)**

## retinoic acid-binding proteins

| CPM | | | | TSM | | | |
|---|---|---|---|---|---|---|---|
| 2acl:A | ~0.0000 | 1xls:C | 22.4501 | 1k74:A | 9.2930 | 1xdk:A | 15.5882 |
| 2acl:C | ~0.0000 | 1xls:A | 22.5071 | 1fm9:A | 9.3501 | 1xdk:E | 15.6471 |
| 2acl:E | ~0.0000 | 1xls:B | 22.5071 | 1fm6:U | 12.7708 | 1fby:A | 16.6471 |
| 2acl:G | ~0.0000 | 1xls:D | 22.5071 | 1fm6:A | 13.4550 | 1fby:B | 20.9928 |
| 1fby:B | 19.7966 | 1fm6:U | 23.3751 | 1xls:A | 15.2137 | 2acl:A | ~100.0000 |
| 1xdk:A | 22.0000 | 1fm6:A | 23.5462 | 1xls:C | 15.2707 | 2acl:C | ~100.0000 |
| 1xdk:E | 22.0000 | 1fm9:A | 25.0855 | 1xls:D | 15.2707 | 2acl:E | ~100.0000 |
| 1fby:A | 22.0985 | 1k74:A | 25.3136 | 1xls:B | 15.3276 | 2acl:G | ~100.0000 |

CPM

| | | | | | |
|---|---|---|---|---|---|
| 1vhc:B | 4.2318 | 1vkb:A | 9.9835 | 1vdh:D | 15.9606 |
| 1sdi:A | 4.3452 | 2a3q:A | 10.1751 | 1vdh:A | 16.1084 |
| 1y7i:A | 4.3655 | 2aca:B | 10.2232 | 1u61:B | 16.5953 |
| 1y7i:B | 4.4843 | 1zn6:A | 10.2689 | 1rlh:A | 17.2563 |
| 1sh8:A | 5.1581 | 1z6n:A | 10.4732 | 1yey:C | 18.3673 |
| 1u5w:H | 5.2221 | 1k3r:A | 10.9609 | 1qya:B | 18.4687 |
| 1yx1:A2 | 5.2580 | 1v6h:A | 10.9890 | 1yey:B | 18.5162 |
| 1u5w:E | 5.3200 | 1k3r:B | 11.0206 | 1nx4:C | 18.8641 |
| 1u5w:G | 6.4822 | 1mwq:B | 11.3253 | 1vim:C | 18.8705 |
| 1kyh:A | 6.7399 | 1mwq:A | 11.5202 | 1nx8:C | 18.9040 |
| 1o65:A | 7.0893 | 1v6h:C | 11.5854 | 1nx4:A | 19.0574 |
| 1xfj:A | 7.1391 | 1rw0:A | 12.3317 | 1njh:A | 19.0813 |
| 1tlj:B | 7.1521 | 1zzm:A | 12.4571 | 1z6m:A | 19.2542 |
| 1wlz:D | 7.2617 | 1v6h:B | 12.5152 | 1y8t:A | 19.4195 |
| 1zkd:A | 7.3617 | 1rw0:B | 12.6010 | 1nx4:B | 19.4280 |
| 1zkd:B | 7.3921 | 1zc6:A | 13.3235 | 1tu1:B | 20.3151 |
| 1yx1:A1 | 7.5184 | 1u7i:B | 13.6452 | 1y8t:B | 20.5441 |
| 1vho:A | 7.5833 | 1vp2:B | 13.7492 | 1u69:A | 20.5832 |
| 1xtl:B | 7.8345 | 1nxz:B | 13.7546 | 1u69:B | 20.5832 |
| 1xtl:D | 7.8345 | 1y0h:B | 13.9355 | 1tu1:A | 20.5931 |
| 1xtl:A | 8.0106 | 1nxz:A | 13.9831 | 2cw5:A | 21.0132 |
| 1xtl:C | 8.4507 | 1zp6:A | 14.0403 | 2apl:A | 21.1735 |
| 1vhs:B | 9.1433 | 1vhy:A | 14.3722 | 1u69:D | 21.6682 |
| 2eui:A | 9.1935 | 1yqe:A | 14.4760 | 1u69:C | 21.6802 |
| 1mw7:A | 9.2593 | 1rz3:A | 14.5003 | 1ze0:A | 21.6814 |
| 2esn:D | 9.2718 | 1vhy:B | 14.5342 | 1y8t:C | 23.0587 |
| 1vhs:A | 9.3577 | 1y0h:A | 14.5546 | 2evr:A | 24.9724 |
| 1nc5:A | 9.4563 | 1u7i:A | 15.3551 | 1wu8:A | 26.3959 |
| 1yll:D | 9.5055 | 1wdt:A | 15.4850 | 1wu8:C | 26.4467 |
| 1iul:A | 9.5903 | 1vdh:B | 15.7635 | 1wu8:B | 26.4975 |
| 1t57:A | 9.7037 | 1rv9:A | 15.8516 | 1yyv:A | 31.2357 |
| 1yll:B | 9.7514 | 1vdh:C | 15.9113 | 1t6s:A | 31.6436 |
| 1t8h:A | 9.7805 | 1vdh:E | 15.9113 | 1t6s:B | 31.6436 |

TSM

| | | | |
|---|---|---|---|
| 1y7m:B | 0.1653 | 1tu1:A | 16.8040 |
| 1y7m:A | 0.2479 | 2evr:A | 17.1271 |
| 1yyv:A | 1.3730 | 1t6s:B | 17.2043 |
| 1ze0:A | 3.1858 | 1t6s:A | 17.2811 |
| 1y8t:A | 5.0660 | 1nx8:C | 17.5465 |
| 1y8t:B | 7.2182 | 1tu1:B | 17.9934 |
| 1njh:A | 8.2450 | 1zp6:A | 18.4466 |
| 1y8t:C | 8.9508 | 1wu8:C | 20.2538 |
| 1yey:B | 11.5880 | 1rv9:A | 20.4609 |
| 1z6m:A | 11.7288 | 1wu8:A | 20.5076 |
| 1y0h:B | 11.7419 | 1wu8:B | 20.5076 |
| 1u61:B | 11.7773 | 1vhs:A | 21.3323 |
| 1yey:C | 12.0621 | 1vhs:B | 21.5815 |
| 1t8h:A | 13.1679 | 1rlh:A | 23.2959 |
| 1y0h:A | 13.9272 | 1rz3:A | 23.7753 |
| 1nx4:B | 14.4970 | 1t57:A | 24.9630 |
| 1k3r:A | 14.7272 | 1zzm:A | 26.7288 |
| 1yqe:A | 15.2753 | 1v6h:B | 27.7035 |
| 1vim:C | 15.4959 | 1v6h:A | 27.9609 |
| 1k3r:B | 15.7163 | 1v6h:C | 28.0488 |
| 1u7i:A | 15.9309 | 1nc5:A | 28.7741 |
| 1nx4:C | 16.0751 | 2cw5:A | 29.1307 |
| 1nx4:A | 16.4959 | 1mwq:A | 30.9976 |
| 1u7i:B | 16.5692 | | |

[NOTE: only structures w/ CPM and TSM values within 10.0 units of those of the training structures are shown]

| | | | | | |
|---|---|---|---|---|---|
| 1k3r | A-32 | H-83 | L-30 | R-33 | AAAA |
| 1k3r | A-32 | H-83 | L-30 | R-33 | BBBB |
| 1mwq | A-26 | H-0 | L-31 | R-27 | AAAA |
| 1nc5 | A-36 | C-48 | L-73 | R-38 | AAAA |
| 1njh | A-11 | H-24 | L-92 | R-15 | AAAA |
| 1nx4 | A-119 | H-251 | L-265 | R-264 | AAAA |
| 1nx4 | A-119 | H-251 | L-265 | R-264 | BBBB |
| 1nx4 | A-119 | H-251 | L-265 | R-264 | CCCC |
| 1nx8 | A-119 | H-251 | L-265 | R-264 | CCCC |
| 1rlh | A-5 | C-94 | L-78 | R-80 | AAAA |
| 1rv9 | A-172 | H-135 | F-173 | R-240 | AAAA |
| 1rz3 | A-99 | H-57 | L-97 | R-22 | AAAA |
| 1t57 | A-45 | C-6 | L-20 | R-47 | AAAA |
| 1t6s | A-131 | H-161 | L-134 | R-132 | AAAA |
| 1t6s | A-131 | H-161 | L-134 | R-132 | BBBB |
| 1t8h | A-25 | C-242 | F-5 | R-14 | AAAA |
| 1tu1 | A-110 | H-7 | L-112 | R-107 | AAAA |
| 1tu1 | A-110 | H-7 | L-112 | R-107 | BBBB |
| 1u61 | A-32 | H-25 | L-58 | R-57 | BBBB |
| 1u7i | A-30 | H-62 | L-56 | R-55 | AAAA |
| 1u7i | A-30 | H-62 | L-56 | R-55 | BBBB |
| 1v6h | A-21 | H-62 | L-27 | R-70 | AAAA |
| 1v6h | A-21 | H-62 | L-27 | R-70 | BBBB |
| 1v6h | A-21 | H-62 | L-27 | R-70 | CCCC |

.)

| | | | | | |
|---|---|---|---|---|---|
| 1vhs | A-104 | C-91 | L-3 | R-106 | AAAA |
| 1vhs | A-104 | C-91 | L-3 | R-106 | BBBB |
| 1vim | A-51 | H--2 | L-56 | R-55 | CCCD* |
| 1wu8 | A-77 | H-36 | F-132 | R-75 | AAAA |
| 1wu8 | A-77 | H-36 | F-132 | R-75 | BBBB |
| 1wu8 | A-77 | H-36 | F-132 | R-75 | CCCC |
| 1y0h | A-24 | C-39 | L-26 | R-23 | AAAA |
| 1y0h | A-24 | C-39 | L-26 | R-23 | BBBB |
| 1y8t | A-137 | H-49 | F-206 | R-140 | AAAA |
| 1y8t | A-137 | H-49 | F-206 | R-140 | BBBB |
| 1y8t | A-137 | H-49 | F-206 | R-140 | CCCC |
| 1yey | A-380 | H-285 | L-320 | R-379 | BBBB |
| 1yey | A-380 | H-285 | L-320 | R-379 | CCCC |
| 1yqe | A-172 | H-76 | L-120 | R-123 | AAAA |
| 1yyv | A-104 | H-28 | L-38 | R-42 | AAAA |
| 1z6m | A-100 | H-81 | F-99 | R-40 | AAAA |
| 1ze0 | A-146 | H-104 | F-147 | R-148 | AAAA |
| 1zp6 | A-145 | C-123 | F-144 | R-113 | AAAA |
| 1zzm | A-195 | H-157 | L-197 | R-237 | AAAA |
| 2cw5 | A-79 | H-38 | F-138 | R-77 | AAAA |
| 2evr | A-39 | C-31 | L-25 | R-42 | AAAA |

**Table 5 (part 4 of 7)**

## heme-bound Nitric Oxide-binding proteins

| CPM | | | | TSM | | | |
|---|---|---|---|---|---|---|---|
| 1zol:B | 19.6326 | | | 1zol:A | 6.7738 | | |
| 1ozw:A | 20.3294 | | | 1xk3:A | 6.9994 | | |
| 1zol:A | 20.6659 | | | 1zol:B | 7.2905 | | |
| 1xk3:A | 20.9409 | | | 1ozw:A | 7.6661 | | |
| 1zbr:A | 2.2095 | * 1qvw:B | 11.2707 | * 2arz:A | 1.7076 | 1xtm:A | 29.9824 |
| 1zbr:B | 2.2095 | * 1xtm:A | 11.2875 | * 1wdt:A | 1.7920 | 1tt4:A | 30.0682 |
| 1xby:A | 5.3211 | * 1qvw:A | 11.3063 | * 2arz:B | 2.1786 | 1qvz:A | 30.1508 |
| 1xbv:B | 5.3922 | * 1xtl:B | 14.3486 | * 1vkh:B | 4.6930 | 1qvv:B | 30.2762 |
| 1xbx:A | 5.3955 | * 1xtl:C | 14.5246 | * 2b4w:A | 5.9513 | 2f9c:B | 30.3874 |
| 1xby:B | 5.4021 | * 1xtl:A | 14.6127 | * 1uan:A | 9.4262 | 1qvz:B | 30.4176 |
| 1xbx:B | 5.4154 | * 1xtl:D | 14.6127 | * 1uan:B | 9.5238 | 1zbs:A | 30.6818 |
| 1xbz:B | 5.5215 | * 1xqb:A | 15.6164 | * 1zsw:A | 10.2977 | 2f9c:A | 32.4889 |
| 1iuk:A | 6.4279 | * 1xqb:B | 15.6849 | * 1xe8:A | 11.8042 | 1xcc:A | 36.9135 |
| 1tt4:A | 6.5303 | * 1o61:B | 17.2366 | * 1xtl:A | 17.0775 | 1xcc:D | 36.9478 |
| 1qy9:A | 7.2523 | * 1o69:B | 17.2473 | * 1xtl:B | 17.0775 | 1xcc:B | 36.9972 |
| 2a67:D | 7.3117 | * 2f9c:A | 18.0137 | * 1xtl:D | 17.0775 | 1xcc:C | 37.4504 |
| 1qy9:D | 7.4473 | * 2f9c:B | 18.6037 | * 1xtl:C | 17.1655 | 1xbx:A | 50.2146 |
| 2a67:A | 7.4492 | * 1zsw:A | 18.7040 | * 1xqb:A | 17.6712 | 1xby:A | 50.7034 |
| 1qy9:B | 7.4775 | * 1xcc:B | 19.4334 | * 1xqb:B | 17.6712 | 1qya:A | 51.1219 |
| 1qya:A | 8.0950 | * 1xcc:A | 19.4460 | 1o69:B | 18.1097 | 1xbx:B | 51.2000 |
| * 1u9c:A | 9.5460 | * 1xcc:C | 19.6034 | 1o61:B | 18.2038 | 1xby:B | 51.3812 |
| * 1qvz:A | 10.3853 | * 1xcc:D | 19.9082 | 2a67:D | 19.3501 | 1xbz:B | 52.4540 |
| * 1qvz:B | 10.4966 | * 1vkh:B | 22.4164 | 2a67:A | 19.4883 | 1xbv:B | 52.5123 |
| * 1zbs:A | 10.5114 | * 1uan:A | 23.8290 | 1u9c:A | 25.9022 | 1qy9:B | 52.6126 |
| * 1qvv:C | 10.8514 | * 1uan:B | 23.8683 | 1qvv:D | 28.3240 | 1qy9:D | 52.8039 |
| * 1qvv:A | 10.8939 | * 2b4w:A | 24.7971 | 1rw7:A | 28.7382 | 1qy9:A | 52.9730 |
| * 1rw7:A | 10.9505 | * 2arz:B | 28.7582 | 1qvv:A | 28.8268 | 1iuk:A | 55.1882 |
| * 1qvv:B | 10.9945 | * 2arz:A | 29.6158 | 1qvw:A | 29.2536 | 1zbr:B | 66.8571 |
| * 1xe8:A | 11.0664 | * 1wdt:A | 30.3272 | 1qvw:B | 29.3370 | 1zbr:A | 68.8762 |
| * 1qvv:D | 11.2290 | | | 1qvv:C | 29.3823 | | |

## heme-NO-binding proteins (cont'd.)

| | | | | | | |
|---|---|---|---|---|---|---|
| | 1o61 | G-350 | C-101 | F-80 | H-336 | BBBB |
| | 1o69 | G-350 | C-101 | F-80 | H-336 | BBBB |
| | 1qvv | G-25 | H-108 | F-17 | Y-13 | AAAA |
| | 1qvv | G-25 | H-108 | F-17 | Y-13 | BBBB |
| | 1qvv | G-25 | H-108 | F-17 | Y-13 | CCCC |
| | 1qvv | G-25 | H-108 | F-17 | Y-13 | DDDD |
| | 1qvw | G-25 | H-108 | F-17 | Y-13 | AAAA |
| | 1qvw | G-25 | H-108 | F-17 | Y-13 | BBBB |
| | 1qvz | G-25 | H-108 | F-17 | Y-13 | AAAA |
| | 1qvz | G-25 | H-108 | F-17 | Y-13 | BBBB |
| | 1rw7 | G-25 | H-108 | F-17 | Y-13 | AAAA |
| | 1u9c | G-22 | H-96 | F-100 | H-12 | AAAA |
| * | 1uan | G-20 | C-18 | F-213 | R-198 | AAAA |
| * | 1uan | G-20 | C-18 | F-213 | R-198 | BBBB |
| * | 1vkh | G-112 | H-36 | F-50 | H-109 | BBBB |
| * | 1wdt | G-89 | H-432 | F-87 | R-63 | AAAA |
| | 1xcc | H-39 | C-73 | F-71 | H-79 | AAAA |
| | 1xcc | H-39 | C-73 | F-71 | H-79 | BBBB |
| | 1xcc | H-39 | C-73 | F-71 | H-79 | CCCC |
| | 1xcc | H-39 | C-73 | F-71 | H-79 | DDDD |
| * | 1xe8 | G-93 | H-202 | F-178 | H-42 | AAAA |
| * | 1xqb | G-34 | H-120 | F-118 | H-13 | AAAA |
| * | 1xqb | G-34 | H-13 | F-118 | H-120 | BBBB |
| * | 1xtl | H-120 | H-86 | F-114 | H-121 | AAAA |
| * | 1xtl | H-120 | H-86 | F-114 | H-121 | BBBB |
| * | 1xtl | H-120 | H-86 | F-114 | H-121 | CCCC |
| * | 1xtl | H-120 | H-86 | F-114 | H-121 | DDDD |
| | 1xtm | G-123 | H-112 | F-114 | H-120 | AAAA |
| | 1zbs | G-100 | C-106 | F-125 | Y-277 | AAAA |
| * | 1zsw | H-46 | H-9 | F-256 | R-253 | AAAA |
| * | 2arz | H-201 | C-40 | F-110 | Y-17 | AAAA |
| * | 2arz | H-201 | C-40 | F-110 | Y-17 | BBBB |
| * | 2b4w | G-191 | C-205 | F-148 | R-136 | AAAA |
| | 2f9c | G-212 | H-213 | F-191 | R-215 | AAAA |
| | 2f9c | G-212 | H-213 | F-191 | R-215 | BBBB |

**Table 5 (part 5 of 7)**

## Unbound Nitric Oxide-binding proteins

| CPM | | TSM | |
|---|---|---|---|
| 1zgn:B | 25.5828 | 1zgn:A | 5.7669 |
| 1zgn:A | 26.2577 | 1zgn:B | 5.8896 |
| 1xa0:A | 1.9108 | * 1uc2:A | 10.9604 |
| 1xa0:B | 2.0976 | * 1vim:B | 13.3798 |
| 1tzz:A | 5.5058 | * 1tzz:B | 15.7609 |
| 1tzz:B | 6.4538 | * 1zn6:A | 16.0147 |
| 1wu8:A | 7.2081 | 1tzz:A | 17.1025 |
| 1vim:B | 13.9373 | 1vbk:A | 19.7980 |
| * 1uc2:A | 14.7380 | 1vbk:B | 20.7846 |
| * 1zn6:A | 15.0978 | 1wu8:A | 62.0812 |
| * 1vbk:B | 20.0751 | 1xa0:B | 93.1935 |
| * 1vbk:A | 21.0505 | 1xa0:A | 94.8195 |

| | | | | | |
|---|---|---|---|---|---|
| 1tzz | Y-2061 | R-2060 | I-2018 | G-2059 | BBBB |
| * 1uc2 | Y-451 | R-408 | I-74 | G-407 | AAAA |
| 1vbk | Y-8 | R-50 | I-116 | G-49 | AAAA |
| 1vbk | Y-8 | R-50 | I-116 | G-49 | BBBB |
| 1vim | Y-47 | R-44 | I-83 | G-46 | BBBB |
| * 1zn6 | Y-156 | R-61 | I-153 | G-60 | AAAA |

**Table 5 (part 6 of 7)**

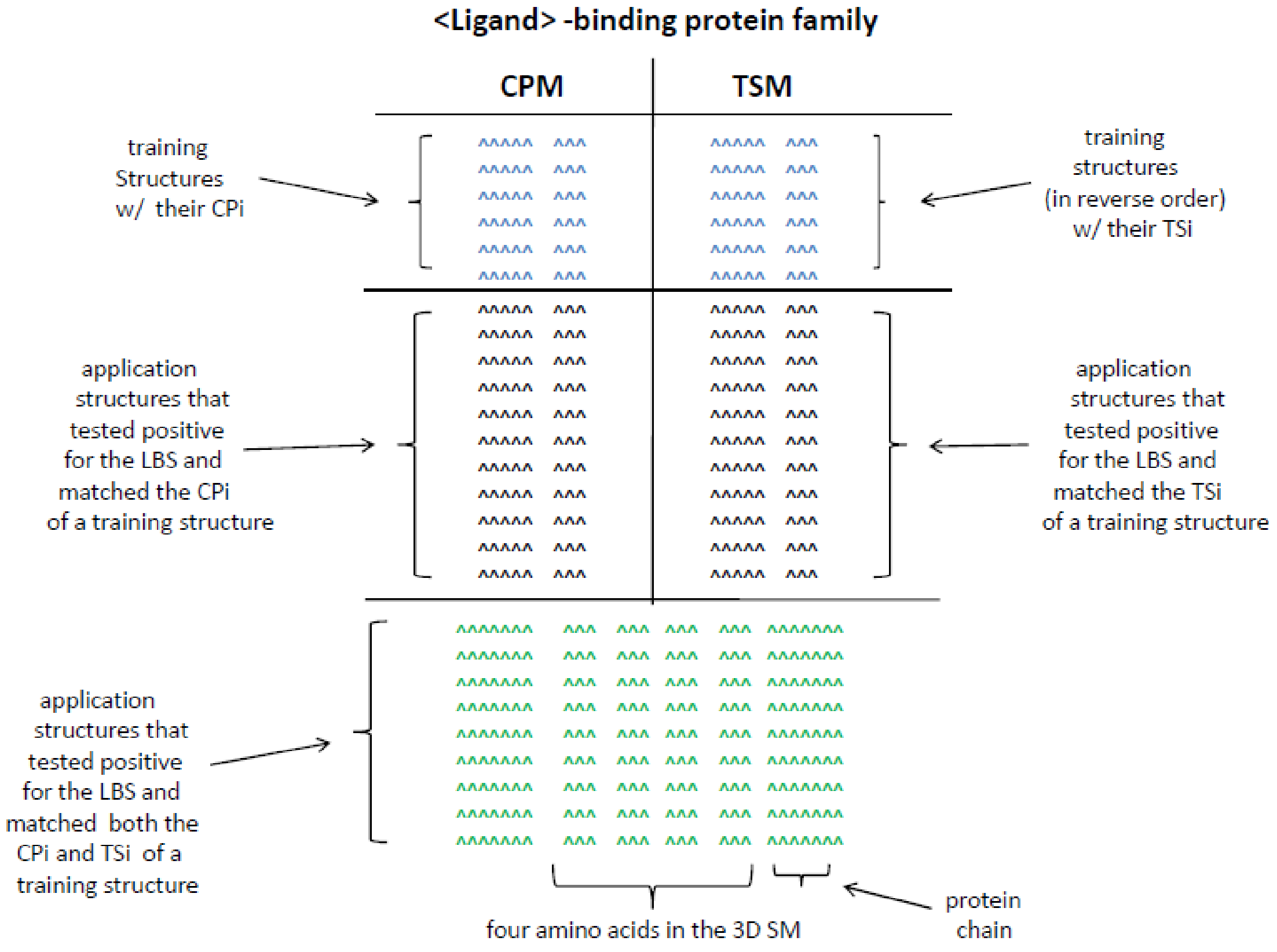

**Table 5 (part 7 of 7)**

| PDB ID | LBS Detected | PDB Header | Source Organism | Remarks | Published Reference |
|---|---|---|---|---|---|
| 1RU8 | GTP[d] | STRUCTURAL GENOMICS, UNKNOWN FUNCTION 11-DEC-03 | *Pyrococcus furiosus* | PUTATIVE N-TYPE ATP PYROPHOSPHATASE; NSG TARGET PFR23 | none |
| 1XTL[a] | GTP | STRUCTURAL GENOMICS, UNKNOWN FUNCTION 22-OCT-04 | *Bacillus subtilis* | P104H MUTANT OF HYPOTHETICAL SUPEROXIDE DISMUTASE-LIKE PROTEIN YOJM | none |
| 1FL9 | ATP[e] | STRUCTURAL GENOMICS, UNKNOWN FUNCTION 13-AUG-00 | *Haemophilus influenzae* | THE YJEE PROTEIN: HYPOTHETICAL PROTEIN HI0065 A NUCLEOTIDE-BINDING FOLD, A PUTATIVE ATPASE INVOLVED IN CELL WALL SYNTHESIS | Proteins. 2002 Aug 1;48(2):220-6 |
| 1MWW | ATP | STRUCTURAL GENOMICS, UNKNOWN FUNCTION 01-OCT-02 | *Haemophilus influenzae* | STRUCTURE OF THE HYPOTHETICAL PROTEIN HI1388.1 WITH A TAUTOMERASE/MIF FOLD | none |
| 1NF2 | ATP | STRUCTURAL GENOMICS/UNKNOWN FUNCTION 12-DEC-02 | *Thermotoga maritima* | PROTEIN TM0651, A PHOSPHATASE with a NEW FOLD AND A UNIQUE SUBSTRATE BINDING DOMAIN | none |
| 1O1Y | ATP | STRUCTURAL GENOMICS, UNKNOWN FUNCTION 12-FEB-03 | *Thermotoga maritima* | CONSERVED HYPOTHETICAL PROTEIN TM1158, A PUTATIVE GLUTAMINE AMIDO TRANSFERASE | Proteins. 2004 Mar 1;54(4):801-5 |
| 1RKI | ATP | STRUCTURAL GENOMICS, UNKNOWN FUNCTION 21-NOV-03 | *Pyrobaculum aerophilum* | PROTEIN PAG5_736 WITH THREE DISULFIDE BONDS | PLoS Biol. 2005 Sep;3(9):e309. Epub 2005 Aug 23 |
| 1RKQ | ATP | STRUCTURAL GENOMICS, UNKNOWN FUNCTION 23-NOV-03 | *Escherichia coli* | NYSGRC TARGET T1436: HYPOTHETICAL PROTEIN YIDA TWO-DOMAIN STRUCTURE W/ BETA-ALPHA SANDWICH; CONTAINS MG++ | none |
| 1T57 | ATP | STRUCTURAL GENOMICS, UNKNOWN FUNCTION 03-MAY-04 | *Methanobacterium thermoautotrophicum* | CONSERVED PROTEIN MTH1675 | none |
| 1TQ8 | ATP | STRUCTURAL GENOMICS, UNKNOWN FUNCTION 16-JUN-04 | *Mycobacterium tuberculosis H37RV* | HYPOTHETICAL PROTEIN RV1636, NYSGRC TARGET T1533 | none |
| 1WM6 | ATP | STRUCTURAL GENOMICS, UNKNOWN FUNCTION 04-JUL-04 | *Thermus thermophilus HB8* | PROTEIN TT0310: A PHENYLACETIC ACID DEGRADATION PROTEIN PAAI; THIOESTERASE W/ HOT DOG FOLD | J Mol Biol. 2005 Sep 9;352(1):212-28 |
| 2CVL | ATP | STRUCTURAL GENOMICS, UNKNOWN FUNCTION 08-JUN-05 | *Thermus thermophilus HB8* | PROTEIN TTHA0137: A PROTEIN TRANSLATION INITIATION INHIBITOR | none |
| 1IUK | sialic acid | STRUCTURAL GENOMICS, UNKNOWN FUNCTION 05-MAR-02 | *Thermus thermophilus* | HYPOTHETICAL PROTEIN TT1466, A CONSERVED COA-BINDING PROTEIN | none |
| 1SQH | sialic acid | STRUCTURAL GENOMICS, UNKNOWN FUNCTION 18-MAR-04 | *Drosophila malonogaster* | HYPOTHETICAL PROTEIN CG14615-PA (Q9VR51), NSGC TARGET FR87 | none |
| 1VKA | sialic acid | STRUCTURAL GENOMICS, UNKNOWN FUNCTION 10-MAY-04 | *Homo sapiens* | HYPOTHETICAL PROTEIN Q15691: N-TERMINAL FRAGMENT MICROTUBULE-ASSOCIATED PROTEIN RP/EB FAMILY; SYN.: APC-BINDING PROTEIN EB1 | none |
| 1Y6Z | sialic acid | STRUCTURAL GENOMICS, UNKNOWN FUNCTION 07-DEC-04 | *Plasmodium falciparum* | C-TERMINAL DOMAIN OF PUTATIVE HEAT SHOCK PROTEIN PF14_0417 (CHAPERONE) | none |
| 1NJH | retinoic acid | STRUCTURAL GENOMICS, UNKNOWN FUNCTION 31-DEC-02 | *Bacillus subtilis* | THE YOJF PROTEIN | none |
| 1NX4 | retinoic acid | UNKNOWN FUNCTION, 08-FEB-03 | *Erwinia carotovora* | CARBAPENEM SYNTHASE (CARC) W/ JELLY ROLL FOLD | J Biol Chem. 2003 Jun 6;278(23):20843-50. Epub 2003 Feb 28 |
| 1NX8 | retinoic acid | UNKNOWN FUNCTION, 10-FEB-03 | *Erwinia carotovora* | CARBAPENEM SYNTHASE (CARC) WITH JELLY ROLL FOLD COMPLEXED WITH N-ACETYL PROLINE | J Biol Chem. 2003 Jun 6;278(23):20843-50. Epub 2003 Feb 28 |
| 1RLH | retinoic acid | STRUCTURAL GENOMICS, UNKNOWN FUNCTION 25-NOV-03 | *Thermoplasma acidophilum* | A CONSERVED HYPOTHETICAL PROTEIN FROM GENE CAC12474; | none |
| 1RV9 | retinoic acid | UNKNOWN FUNCTION, 13-DEC-03 | *Neisseria meningitidis* | A CONSERVED HYPOTHETICAL PROTEIN NMB0706 W/ ALPHA-BETA-BETA-ALPHA STRUCTURE | none |
| 1TU1 | retinoic acid | STRUCTURAL GENOMICS, UNKNOWN FUNCTION 24-JUN-04 | *Pseudomonas aeruginosa* | HYPOTHETICAL PROTEIN PA0094 | none |
| 1U6L | retinoic acid | UNKNOWN FUNCTION, 30-JUL-04 | *Pseudomonas aeruginosa* | HYPOTHETICAL PROTEIN | none |
| 1VIM | retinoic acid | STRUCTURAL GENOMICS, UNKNOWN FUNCTION 01-DEC-03 | *Archaeoglobus fulgidus* | HYPOTHETICAL PROTEIN AF1796 | J Neurochem. 2000 Oct;75(4):1475-86 |
| 1WU8 | retinoic acid | STRUCTURAL GENOMICS, UNKNOWN FUNCTION 02-DEC-04 | *Pyrococcus horikoshii OT3* | HYPOTHETICAL PROTEIN PH0452 | none |
| 1Y8T | retinoic acid | UNKNOWN FUNCTION, 13-DEC-04 | *Mycobacterium tuberculosis* | PROTEOLYTICALLY ACTIVE FORM OF HYPOTHETICAL PROTEIN RV0983, A SERINE PROTEASE-HTRA HOMOLOG | none |
| 1YEY | retinoic acid | STRUCTURAL GENOMICS, UNKNOWN FUNCTION 28-DEC-04 | *Xanthomonas campestris PV. Str. ATCC 33913* | L-FUCONATE DEHYDRATASE | none |
| 1Z6M | retinoic acid | STRUCTURAL GENOMICS, UNKNOWN FUNCTION 22-MAR-05 | *Enterococcus faecalis V583* | CONSERVED HYPOTHETICAL PROTEIN | none |
| 2EVR | retinoic acid | STRUCTURAL GENOMICS, UNKNOWN FUNCTION 31-OCT-05 | *Nostoc punctiforme PCC 73102* | PROTEIN 53698717, CELL WALL-ASSOCIATED HYDROLASES (INVASION-ASSOCIATED PROTEINS) | none |
| 1UAN | heme-NO | STRUCTURAL GENOMICS, UNKNOWN FUNCTION 12-MAR-03 | *Thermus thermophilus HB8* | CONSERVED HYPOTHETICAL PROTEIN TT1542 | none |
| 1VKH | heme-NO | STRUCTURAL GENOMICS, UNKNOWN FUNCTION 20-MAY-04 | *Saccharomyces cerevisiae* | PUTATIVE SERINE HYDROLASE WITH ALPHA/BETA FOLD(YDR428C) | Proteins. 2005 Feb 15;58(3):755-8 |
| 1WDT[a] | heme-NO | STRUCTURAL GENOMICS, UNKNOWN FUNCTION 17-MAY-04 | *Thermus thermophilus HB8* | ELONGATION FACTOR G HOMOLOG TTHG03000958, GTP COMPLEX | none |
| 1XE8 | heme-NO | STRUCTURAL GENOMICS, UNKNOWN FUNCTION 09-SEP-04 | *Saccharomyces cerevisiae* | HYPOTHETICAL 22.5 KD PROTEIN YML079W; NEW SEQUENCE FAMILY OF THE JELLY ROLL FOLD; IN TUB1-CPR3 INTERGENIC REGION; SYN.: YML079WP; CUPIN SUPERFAMILY | Protein Sci. 2005 Jan;14(1):209-15 |
| 1XQB | heme-NO | STRUCTURAL GENOMICS, UNKNOWN FUNCTION 11-OCT-04 | *Haemophilus influenzae* | NESGC TARGET IR47; HYPOTHETICAL UPF0066 PROTEIN HI0510; YAEB PROTEIN | none |
| 1XTL* | heme-NO | STRUCTURAL GENOMICS, UNKNOWN FUNCTION 22-OCT-04 | *Bacillus subtilis* | P104H MUTANT OF S.O.D.-LIKE HYPOTHETICAL PROTEIN YOJM; CU-ZN SOD | none |
| 1ZSW[c] | heme-NO | STRUCTURAL GENOMICS, UNKNOWN FUNCTION 25-MAY-05 | *Bacillus cereus* | GLYOXALASE FAMILY METALLOPROTEIN | none |
| 2ARZ | heme-NO | UNKNOWN FUNCTION, 22-AUG-05 | *Pseudomonas aeruginosa* | HYPOTHETICAL PROTEIN PA4388 | none |
| 2B4W | heme-NO | STRUCTURAL GENOMICS, UNKNOWN FUNCTION 28-SEP-05 | *Leishmania major (protozoa)* | CONSERVED HYPOTHETICAL PROTEIN | none |
| 1UC2 | unbound NO | STRUCTURAL GENOMICS, UNKNOWN FUNCTION 09-APR-03 | *Pyrococcus horikoshii (archaea)* | HYPOTHETICAL EXTEIN; RTCB HOMOLOG PROTEIN OF PH1602 | none |
| 1ZN6 | unbound NO | STRUCTURAL GENOMICS, UNKNOWN FUNCTION 11-MAY-05 | *Bordetella bronchiseptica* | PHAGE-RELATED CONSERVED HYPOTHETICAL PROTEIN GTWLM5; NESGC TARGET BOR19 W/ NOVEL STRUCTURE | none |

[a] A heme-NO (nitric oxide) 3D SM has also been detected in this structure; see 1XTL* below.

[b] A 3D SM corresponding to monomer 2 in the interleukin-2 homodimer complex (3INK:B) was also detected in this structure (see Reyes, V.M., 2008c).

[c] A 3D SM corresponding to monomer 1 in the RAP•RAFRBD complex (RAP•Gmppnp complexed w/ c-RAF1 Ras-binding domain; 1C1Y:A) was also detected in this structure (see Reyes, V.M., 2008c).

[d] The specific GTP LBS detected is the 3D SM corresponding to that in the small, Ras-type G-protein family.

[e] The specific ATP LBS detected is the 3D SM corresponding to that in the ser/thr protein kinase family.

**Table 6.**